\documentclass[prb,twocolumn,superscriptaddress,floatfix]{revtex4}
\usepackage{amsmath}
\usepackage[dvips]{graphicx}
\usepackage[dvips]{color}

\newcommand{\abs}[1]{\ensuremath{\left|#1\right|}}

\begin{document}

\title{Dephasing by a nonstationary classical intermittent noise}

\author{J. Schriefl}
\affiliation{CNRS-Laboratoire de Physique de l'Ecole Normale Sup{\'e}rieure de Lyon, 
46, All{\'e}e d'Italie, 69007 Lyon, France} \affiliation{Institut f{\"u}r Theoretische
Festk{\"o}rperphysik Universit{\"a}t Karlsruhe,
 76128 Karlsruhe, Germany}
\author{M. Clusel}
\affiliation{CNRS-Laboratoire de Physique de l'Ecole Normale Sup{\'e}rieure de Lyon, 
46, All{\'e}e d'Italie, 69007 Lyon, France}
\author{D. Carpentier}
\affiliation{CNRS-Laboratoire de Physique de l'Ecole Normale Sup{\'e}rieure de Lyon, 
46, All{\'e}e d'Italie, 69007 Lyon, France}
\author{P. Degiovanni}
\affiliation{CNRS-Laboratoire de Physique de l'Ecole Normale Sup{\'e}rieure de Lyon, 
46, All{\'e}e d'Italie, 69007 Lyon, France}

\date{\today}

\begin{abstract}
We consider a new phenomenological model for a $1/f^{\mu}$ classical intermittent noise and study its 
effects on the dephasing of a two-level system.  Within this model, the evolution of the relative phase between the 
$|\pm\rangle$ states is described as a continuous time random walk (CTRW). Using renewal theory,  we find exact 
expressions for the dephasing factor and identify the physically relevant various regimes in terms 
of the coupling to the noise. 
In particular, we point out the consequences of the non-stationarity and pronounced non-Gaussian
features of this noise, including some new anomalous and aging dephasing scenarii. 
\end{abstract}

\maketitle

\section{Introduction}

 Recent experimental progress in the study of solid-state quantum bits (Josephson qubits)\cite{Makhlin2001} has stressed the 
 importance of low-frequency noise in the dephasing or decoherence of these two-level 
 systems\cite{Nakamura2002,Vion2002,Simmonds2004}. It now appears 
 that the coupling to low-frequency noise is the main limitation in obtaining long lived phase coherent 
 states of qubits necessary for quantum computation. However, a complete understanding
 of the microscopic origin of $1/f$ noise in solid state physics is not available yet\cite{Weissman1988} 
 and therefore, theoretical studies of the dephasing by such a noise are based on phenomenological 
 models. In the spin-boson model, 
 the environment of the qubit is modeled by a set of harmonic oscillators, with an adequate frequency 
spectrum\cite{Makhlin2004-1,Makhlin2004-2}. Another commonly used
model for a low-frequency noise consists in 
considering the contributions from many independent bistable fluctuators\cite{Paladino2002,Galperin2003}. 
In the semi-classical limit, the noise from each fluctuator is approximated by a telegraph noise of characteristic
switching rate $\gamma$. For a broad distribution  $\sim 1/\gamma$ of switching rates $\gamma$, a $1/f$ spectrum is recovered 
when summing contributions of all fluctuators. 
Such a model is based on observations of telegraph like fluctuations in nano-scale
devices\cite{Buehler2004,Zimmerli1992}, 
but a precise characterization and justification of the broad distribution of switching rates is still 
lacking although the localization of these fluctuators\cite{Farmer1987,Zorin1996} as well as 
their collective or individual nature\cite{Welland1986} have been investigated for a long time. 

 In this paper, we consider a new phenomenological model for the classical low-frequency noise. This model can be 
viewed as the intermittent limit of the sum of telegraphic signals. In this limit, the duration of each plateau of the 
telegraphic signal is assumed to be much shorter than the waiting time between plateaus\cite{Buehler2004}. 
A $1/f$ power spectrum for the intermittent 
noise is then recovered for a distribution of waiting times $\tau$ behaving as $\tau^{-2}$ for large times. Because the average
waiting time is infinite, no time scale characterizes the evolution of the noise which is nonstationary. 
The purpose of this article is to study the effects of such a
low-frequency intermittent noise on the dephasing of a two-level system in order to
identify possible signatures of intermittence. 
 
As we will show, in this model the relative phase $\Phi$ between the states of the qubit 
performs a continuous time random walk\cite{Montroll1965,Haus1987} (CTRW) as time goes on.  Such a CTRW
was considered in the context of $1/f$ current noise by Tunaley\cite{Tunaley1976}, extending the previous 
work of Montroll and Scher on electronic transport\cite{Montroll1973}. 
However, in the present paper it is the integral of noise, and not the noise itself, which performs a CTRW. 
Moreover to our knowledge, 
the precise consequences of CTRWs non-stationarity on dephasing have not been studied. On the 
other hand aging CTRW 
were previously considered in the context of trap models in glassy materials\cite{Monthus1996} and in the study of 
fluorescence of single nanocrystals\cite{Barkai2002,Barkai2003b}. 
Technically, the dephasing factor that
we will consider corresponds to the average Fourier transform of the positional correlation function of the random walk.
 Some of the asymptotic behaviors of this correlation function were already obtained in ref.\onlinecite{Monthus1996}. 
 However, in the present paper we will extend these results to all possible regimes and 
we will present all of these results in a unified framework. The use of renewal theory\cite{Feller1962} greatly enlightens
the origin of non stationarity and enables us to interpret some features of the dephasing scenarii.  

 This paper is organized as follows : in section \ref{sec:Model}, we present 
 our model for the noise and define 
 the quantity of interest, {\it i.e} the dephasing factor of a two-level system coupled to this noise. 
 In section \ref{sec:RandomWalk}, the exact expression for the single Laplace transform of 
 the dephasing factor will be derived and, from this result, the physically relevant weak and strong coupling regimes are
 identified. Moreover, we clarify the origin of non-stationarity and show the relation of our
 problem to renewal theory. For completeness and pedagogy, the effects of standard anomalous diffusion 
 of the phase and of randomness of waiting times on dephasing are compared showing the importance of
 intermittence in the non-stationarity properties of the dephasing scenarii.
 In sections  \ref{sec:symmetric} and \ref{sec:Asymmetric}, we present a complete study of the behavior of the 
 dephasing factor respectively for a noise with a vanishing average amplitude (symmetric noise) and with a finite 
 average one (asymmetric noise). The general discussion of the results is postponed to section \ref{sec:discussion}.  

\section{The model}
\label{sec:Model}

\subsection{Pure dephasing by an intermittent noise}
\label{sec:PureDephasing}

 In this paper, we consider a quantum bit defined as a two-level system with 
 controllable energy difference $\hbar \omega_{0}$ and tunneling amplitude $\Delta$ 
 between the two states $|-\rangle $ and $|+\rangle $ (eigenstates of $\sigma_{z}$).  
 The effect of the environment on this two-level system will 
 be accounted for by a fluctuating 
 shift $\hbar X$ of the energy difference $\hbar \omega_{0}$. Thus the 
 Hamiltonian describing this model is written as 
\begin{equation}\label{eq:Hamiltonian} 
{\cal H} = \frac{\hbar}{2} 
\left( \omega_{0} \: \sigma_z + \Delta\: \sigma_x   -  X \: \sigma_z  \right) \; ,
\end{equation}
 In this paper, we will mainly focus on the case of pure dephasing ($\Delta = 0$). However, 
 as explained below in section \ref{sec:decoherence}, our discussion will also apply 
 to other operating points ($\Delta\neq 0$), including the special points where a careful
 choice of control parameters considerably lower the qubit sensitivity to low frequency 
 noise\cite{Vion2002}.

 Here, we will focus on the effects of a low-frequency classical noise on the qubit. The noise is represented by
 a classical stochastic function corresponding to the fluctuations of the noise in a given sample. 
Within this statistical approach, we focus on the statistical properties ({\it e.g.} the average) of physical quantities
associated with the qubit such as the so-called  ({\it average}) dephasing factor. As we shall see now, its
meaning can be understood by  considering a typical 
Ramsey (interference) experiment on the qubit\cite{Ramsey1950}. 
 
In such an experiment,  the qubit is prepared at initial time $t_{p}$ in a
superposition of the eigenstates of $\sigma_z$,
{\it e.g} $|+\rangle = (|\!\!\uparrow\rangle + |\!\!\downarrow\rangle)/\sqrt{2}$.
Note that throughout this
paper, $t=0$ will correspond to the origin of time for the noise ({\it e.g} the time at which the sample reached the experiment's temperature).
At some later time $t_{p}+\tau_{exp}> t_{p}$, we consider the projection of the evolved qubit state on $|\!\!\uparrow\rangle$.
In the mean time, the state has evolved under Hamiltonian \eqref{eq:Hamiltonian} ($\Delta = 0$)
and both states $|\!\!\uparrow\rangle$ and $|\!\!\downarrow\rangle$ 
have accumulated a random relative phase $\Phi(t_{p},\tau_{exp})$ defined by
\begin{equation}
\label{eq:def-phase}
\Phi(t_{p},\tau_{exp})= \int_{t_{p}}^{t_{p}+\tau_{exp}}X(t)\,dt.
\end{equation}
For a given accumulated phase $\Phi = \Phi(t_{p},\tau_{exp})$,  the quantum probability
$P_{\Phi,\tau_{exp}}(|\!\!\uparrow \rangle)$ to find the qubit in state
$|+\rangle$ at time $t_{p}+\tau_{exp}$ is given by
\begin{equation}
\label{eq:quantum-proba}
P_{\Phi,\tau_{exp}}(|+\rangle)
=\frac{1}{2}
\left[1+\cos{\left(\omega_{0} \tau_{exp}-\Phi(t_{p},\tau_{exp})
\right)}\right]
\end{equation}
 Note that in a given sample, $P_{\Phi,\tau_{exp}}(|+\rangle)$ oscillates between $0$ and
 $1$ as a function of $\tau_{exp}$ (although possibly nonperiodically). The experimental determination of the 
 probability for finding the qubit in the $|+\rangle$ state at time $t_{p}+t$ usually requires
many experimental runs of same duration $\tau_{exp}$.
 The phase fluctuations between different runs induce an attenuation of the amplitudes of these oscillations
  (analogously  to destructive interference effects in optics). Using Bayes theorem,
  the corresponding statistical frequency to find
  the qubit in state $|+\rangle$ after a duration $\tau_{exp}$ is given by the probability
  \begin{align}
  \nonumber
  P_{t_{p},\tau_{exp}} (|+\rangle) & =
  \int d\Phi ~
  P_{\Phi,\tau_{exp}}(|+\rangle) \mathcal{P}(\Phi=\Phi(t_{p},\tau_{exp})) \\
  & =  \frac12
  \left( 1 + \Re \left( D_{t_{p}}(\tau_{exp}) e^{-i\omega_{0} \tau_{exp}} \right)   \right)
  \label{eq:stat-proba}
  \end{align}
  In this expression, the decay rate of these oscillations is encoded in the dephasing factor $D_{t_{p}}(\tau_{exp}) $
   defined as
\begin{equation}
\label{eq:defD}
D_{t_{p}}(\tau_{exp}) =
\overline{\exp{\left(i \Phi(t_{p},\tau_{exp}) \right)}}.
\end{equation}
  In this formula (and only here), the overline denotes an average over all possible configurations of noise $X(t)$
during the experiment.
   
 Note that in deriving eq.(\ref{eq:stat-proba}), statistical independence of the phases $\Phi$ between 
 different runs has been assumed. 
 This is not necessarily true for successive runs in a given sample as correlations of the noise
might lead to a dependence of the distribution of the 
 phase $\Phi(t_{p},\tau_{exp})$ on the starting date $t_{p}$ of the run. Hence throughout this paper, 
 for self-consistency, 
 we will keep track of this effect through a possible $t_{p}$ dependence of the dephasing factor $D_{t_{p}}(\tau_{exp}) $. Its 
 possible implications will be discussed together with our results in section \ref{sec:discussion}.
   
\subsection{A model for classical intermittent noise} 
\label{sec:IntermittentNoise}
   
\begin{figure*}[htbp]
\begin{center}
\begin{picture}(0,0)%
\includegraphics{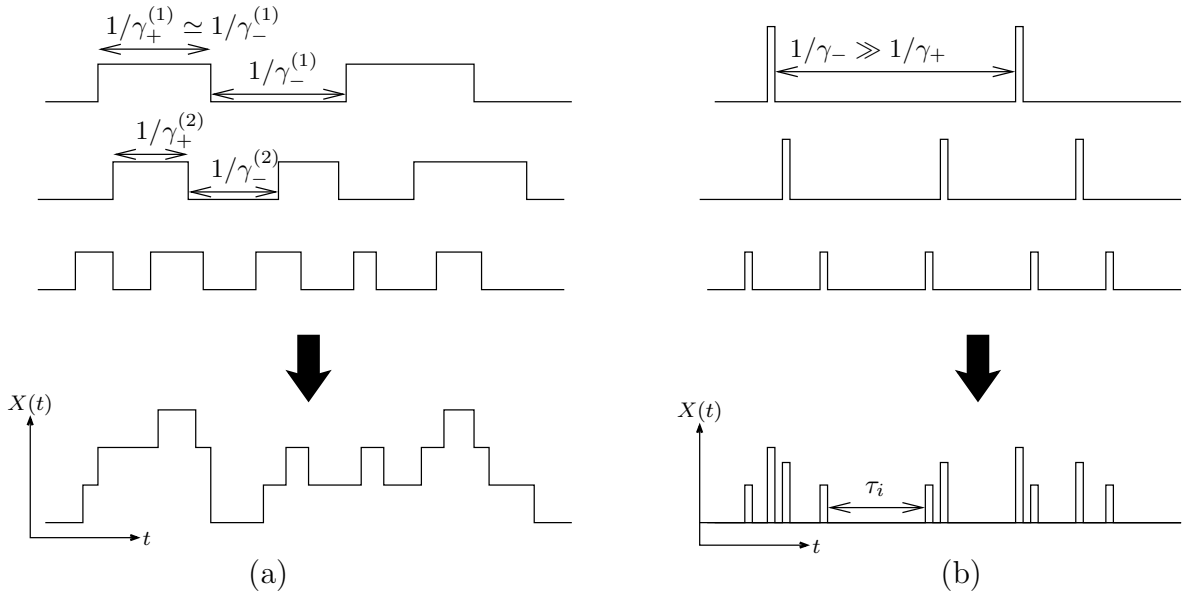}%
\end{picture}%
\setlength{\unitlength}{4144sp}%
\begingroup\makeatletter\ifx\SetFigFont\undefined%
\gdef\SetFigFont#1#2#3#4#5{%
  \reset@font\fontsize{#1}{#2pt}%
  \fontfamily{#3}\fontseries{#4}\fontshape{#5}%
  \selectfont}%
\fi\endgroup%
\begin{picture}(7032,3471)(181,-2320)
\put(5311,-1726){\makebox(0,0)[lb]{\smash{{\SetFigFont{10}{12.0}{\rmdefault}{\mddefault}{\updefault}{\color[rgb]{0,0,0}$\tau_i$}%
}}}}
\put(181,-1231){\makebox(0,0)[lb]{\smash{{\SetFigFont{8}{9.6}{\familydefault}{\mddefault}{\updefault}{\color[rgb]{0,0,0}$X(t)$}%
}}}}
\put(991,-2041){\makebox(0,0)[lb]{\smash{{\SetFigFont{8}{9.6}{\familydefault}{\mddefault}{\updefault}{\color[rgb]{0,0,0}$t$}%
}}}}
\put(946,389){\makebox(0,0)[lb]{\smash{{\SetFigFont{10}{12.0}{\rmdefault}{\mddefault}{\updefault}{\color[rgb]{0,0,0}$1/\gamma_+^{(2)}$}%
}}}}
\put(1396,164){\makebox(0,0)[lb]{\smash{{\SetFigFont{10}{12.0}{\rmdefault}{\mddefault}{\updefault}{\color[rgb]{0,0,0}$1/\gamma_-^{(2)}$}%
}}}}
\put(4186,-1276){\makebox(0,0)[lb]{\smash{{\SetFigFont{8}{9.6}{\familydefault}{\mddefault}{\updefault}{\color[rgb]{0,0,0}$X(t)$}%
}}}}
\put(4996,-2086){\makebox(0,0)[lb]{\smash{{\SetFigFont{8}{9.6}{\familydefault}{\mddefault}{\updefault}{\color[rgb]{0,0,0}$t$}%
}}}}
\put(4861,884){\makebox(0,0)[lb]{\smash{{\SetFigFont{10}{12.0}{\rmdefault}{\mddefault}{\updefault}{\color[rgb]{0,0,0}$1/\gamma_- \gg 1/\gamma_+$}%
}}}}
\put(1621,-2266){\makebox(0,0)[lb]{\smash{{\SetFigFont{12}{14.4}{\rmdefault}{\mddefault}{\updefault}{\color[rgb]{0,0,0}(a)}%
}}}}
\put(5761,-2266){\makebox(0,0)[lb]{\smash{{\SetFigFont{12}{14.4}{\rmdefault}{\mddefault}{\updefault}{\color[rgb]{0,0,0}(b)}%
}}}}
\put(1621,749){\makebox(0,0)[lb]{\smash{{\SetFigFont{10}{12.0}{\rmdefault}{\mddefault}{\updefault}{\color[rgb]{0,0,0}$1/\gamma_-^{(1)}$}%
}}}}
\put(766,1019){\makebox(0,0)[lb]{\smash{{\SetFigFont{10}{12.0}{\rmdefault}{\mddefault}{\updefault}{\color[rgb]{0,0,0}$1/\gamma_+^{(1)}\simeq 1/\gamma_-^{(1)}$}%
}}}}
\end{picture}%
\caption{
\label{fig:DuttaHorn}
Representation of a low-frequency noise as a sum of contributions from telegraphic signals (a). In this first case, 
the switching rates for the up ($\gamma_{+}$) and down ($\gamma_{-}$) states are comparable, and a $1/f$ spectrum 
is recovered for a distribution of switching rates $\sim 1/\gamma$. The intermittent limit (b) corresponds to the limit 
where the noise stays in the down states most of the time ($\gamma_{+} \gg \gamma_{-}$). In this paper, we will approximate 
this intermittent noise by a spike field. For this intermittent noise, a $1/f$ spectrum implies a non stationarity whose 
consequences on dephasing are studied in this paper. 
}
\end{center}
\end{figure*}

  In several experimental situations, the low-frequency noise acting on the qubit is supposed to be due to contributions 
from background charges in the substrate\cite{Paladino2002,Galperin2003}. 
When the dephasing is dominated by the low-frequency fluctuators, a 
semi-classical approach, in which the noise is modeled by a classical field, appears sufficient\cite{Paladino2002,Galperin2003,Falci2003}. 
 In this case, the noise  is described by a Dutta-Horn model\cite{Dutta1981}. In its simplest form, the potential
$X(t)$ is written as the sum of the contributions of many telegraphic signals, each with a characteristic switching 
rate $\gamma$ between the up and down states (see figure \ref{fig:DuttaHorn}a).  
For switching rates distributed according to an algebraic distribution 
$p(\gamma) \simeq 1/\gamma$, the power spectrum of the corresponding noise has a $1/f$ low-frequency behavior. 
 
   In this paper, motivated by several noise signatures\cite{Zimmerli1992,Zorin1996}, 
we will focus on the intermittent limit for such a Dutta-Horn model, and propose a phenomenological 
spike field to study its effects on qubit dephasing. By intermittence, we mean for the case of the telegraphic noise that the 
switching rate $\gamma_{+}$ from the up to the down states is much larger than the switching rate 
$\gamma_{-}$ from the down to the up states (or vice-versa). In this limit, the total noise reduces to a collection 
of well defined events, separated by waiting times $\tau_{i}$ (see figure \ref{fig:DuttaHorn}b). If considered on times 
much longer than the typical duration $\tau_{0}$ of these events 
(or at frequencies smaller than $1/\tau_{0}$), we can approximate 
this noise by a spike field consisting of a succession of delta functions of weight $x_{i}$
corresponding to the 
integral over time of the corresponding events of the intermittent field (Fig.\ref{fig:DuttaHorn}b). 

\begin{figure}[htbp]
\begin{center}
\begin{picture}(0,0)%
\includegraphics{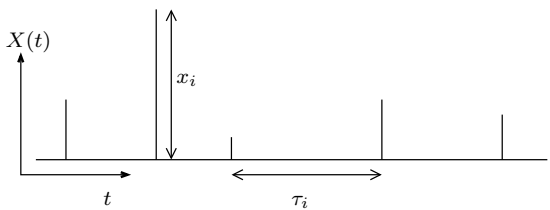}%
\end{picture}%
\setlength{\unitlength}{4144sp}%
\begingroup\makeatletter\ifx\SetFigFont\undefined%
\gdef\SetFigFont#1#2#3#4#5{%
  \reset@font\fontsize{#1}{#2pt}%
  \fontfamily{#3}\fontseries{#4}\fontshape{#5}%
  \selectfont}%
\fi\endgroup%
\begin{picture}(3252,1213)(1,-3062)
\put(1711,-3031){\makebox(0,0)[lb]{\smash{{\SetFigFont{8}{9.6}{\familydefault}{\mddefault}{\updefault}{\color[rgb]{0,0,0}$\tau_{i}$}%
}}}}
\put(  1,-2086){\makebox(0,0)[lb]{\smash{{\SetFigFont{8}{9.6}{\familydefault}{\mddefault}{\updefault}{\color[rgb]{0,0,0}$X(t)$}%
}}}}
\put(586,-3031){\makebox(0,0)[lb]{\smash{{\SetFigFont{8}{9.6}{\familydefault}{\mddefault}{\updefault}{\color[rgb]{0,0,0}$t$}%
}}}}
\put(1022,-2301){\makebox(0,0)[lb]{\smash{{\SetFigFont{8}{9.6}{\familydefault}{\mddefault}{\updefault}{\color[rgb]{0,0,0}$x_i$}%
}}}}
\end{picture}%
\caption{Representation of the random spike field used to model the
intermittent low-frequency noise in this work. This noise is described
by the distributions of the phase pulses $x_{i}$ and of the
waiting time intervals $\tau_{i}$ between the spikes.}
\label{fig:spikefield}
\end{center}
\end{figure}

 More precisely, denoting by $t=0$ the origin of time, and by $\tau_{i}$ the successive waiting times between the 
 spikes (or events), we know that the $n$th spike occurs at time $t_{n}=\sum_{i=1,n}\tau_{i}$. The  value of the 
stochastic intermittent classical noise $X(t)$ (see Fig.~\ref{fig:spikefield}) is then 
\begin{equation}
\label{eq:defnoise}
X(t)=\sum_i x_{i}\,\delta(t-t_i)
\end{equation}

 In the following, we will consider the dephasing produced by this spike field. We expect any short time details 
like the specific shape of the real pulses to be irrelevant in the limit of typical dephasing time long compared to $\tau_{0}$. 
 Within this approximation, a noise signal is fully characterized by the collection of waiting times $\tau_{i}$ and pulse 
 amplitudes $x_{i}$ that occur as time goes on. We will assume these two quantities to be independent from each other, 
 and completely uncorrelated in time. We will then characterize such a noise solely by two independent probability 
 distributions $\psi(\tau)$ and $p(x)$ for the $\tau_{i}$ and $x_{i}$ respectively. 
 
  The distribution of pulse amplitudes $p(x)$ will be assumed to have at least its two first moments finite, and denoted in the 
following by
\begin{subequations}\label{eq:moments-p}  
\begin{align}
h &= \overline{x} = \int_{-\infty}^{+\infty} xp(x)\,dx\,,\\
g  &= \sqrt{\overline{x^2}} =\left( \int_{-\infty}^{+\infty} x^2 p(x)\,dx\right)^{1/2}\,.
\end{align}
\end{subequations}
 We will consider separately the case of zero average (symmetric noise)  and of non-zero average
 (asymmetric noise) since the latter induces
specific features of the dephasing factor, discussed in section \ref{sec:Asymmetric}. 

 We will consider algebraic distribution of waiting times $\psi(\tau)$, parametrized by a single parameter 
 $\mu$ :  
\begin{equation}
\label{def-P}
\psi(\tau)=\frac{\mu}{\tau_{0}}\left(\frac{\tau_{0}}{\tau_{0}+\tau} \right)^{1+\mu}
\; ,\ \tau>0\,.
\end{equation}
As we will show in section \ref{sec:SpectralProperties}, 
a $1/f^{\mu}$ power spectrum for this intermittent spike field follows naturally from such a choice for 
$\psi(\tau)$. As we shall see in this paper, the above algebraic distribution of waiting times 
allows to correctly capture the essential features of the dephasing scenarii generated by an intermittent noise. 
The first and second moments of $\psi$ are finite respectively
 for $\mu>1$ and $\mu>2$ and are given by
 \begin{equation}\label{eq:def-moments}
\left< \tau  \right> =\frac{\tau_{0}}{\mu -1} \qquad ;\qquad
\left< \tau^{2} \right>= \frac{2\tau_{0}^{2}}{(\mu -1)(\mu -2)}
\end{equation}
Let us note that the above defined pulse noise contains two independent potential sources of dephasing: 
The randomness of the pulse weights and the randomness the waiting times. Their respective effects will be
compared in section \ref{sec:RandomWalk}.

\subsection{Decoherence at optimal points}
\label{sec:decoherence}

 Before turning to the detailed study of pure dephasing ($\Delta=0$ in \eqref{eq:Hamiltonian}), let us mention 
that our discussion can be easily extended to the study of dephasing in the presence of a 
transverse coupling in \eqref{eq:Hamiltonian}, in particular at the so-called optimal points. 
They correspond to configurations where
the fluctuations of the effective qubit level splitting are only quadratic
in the noise amplitude. The qubit can be operated at these optimal points by
a careful choice of the control parameters  $\omega_0$ and $\Delta$ 
of the qubit and then, the influence of low frequency noise
can be reduced considerably\cite{Vion2002}. 
For the Hamiltonian considered in the present work \eqref{eq:Hamiltonian}
such an optimal point is reached for transverse coupling to the noise 
($\omega_0=0$ and $\Delta\neq 0$). In this case and assuming 
the amplitude of the noise to be small compared to 
the control parameter $\Delta$, the effective qubit 
level splitting is given by 
$\sqrt{\Delta^2+X(t)^2} \approx \Delta + X(t)^2/(2\Delta)$. 
Hence, the dephasing effect of a linear transverse noise can be accounted for using an effective quadratic longitudinal
noise. The corresponding dephasing factor is then given by \eqref{eq:defD} and \eqref{eq:def-phase} with 
the replacement $X \to X^2/(2\Delta)$.

In addition, the transverse noise at an optimal point induces transitions
between the eigenstates of the qubit, {\it i.e.} it leads to relaxation.
The Fermi Golden Rule relaxation rate $\Gamma_{r}$, which is used for estimating the
relaxation contribution to dephasing, involves the power spectrum of the noise
at the resonance frequency of the qubit\cite{Makhlin2001}. The total dephasing
rate is obtained by summing the contribution of relaxation given by $\Gamma_{r}/2$
and the contribution of pure dephasing due to the above effective longitudinal noise.

In general the statistics of $X$ and $X^2$ are very different and a special 
treatment is needed to derive the dephasing factor at optimal points 
\cite{Makhlin2004-1}. However, for the noise considered in the
present work, the effective quadratically coupled 
longitudinal noise can be viewed as en effective linearly coupled noise of the same type
but with renormalized parameters. The renormalized distribution 
of the spike intensities is now given by ($x\geq 0$): 
\begin{equation}
\tilde p(x) = \sqrt{\frac{\Delta\tau_{0}}{8x}}\,\lbrace p\left(\sqrt{2x\Delta\tau_{0}}\right)
+p\left(-\sqrt{2x\Delta}\right)\rbrace\,,
\end{equation}
where $\tau_{0}$ is a microscopic time scale needed to regularize the square of delta functions.
Therefore, our results for the longitudinal noise presented in section \ref{sec:PureDephasing} can
also be used to describe the effect of a transverse noise.

\subsection{Spectral properties of the intermittent noise}
\label{sec:SpectralProperties}

\subsubsection{One and two point correlation functions}

To make contact with other descriptions of low-frequency noise, we
will determine the behavior of the two time correlation function of
our noise, or equivalently of its spectral density. However, as we
will see, this spectral density is far from enough to characterize the
statistics of the noise for small $\mu$, in particular due to its non stationarity.
 We will consider for simplicity the case of a non zero average $h=\overline{x}$. 

\paragraph{Time dependent average}
 Let us consider the average of the noise amplitude $X(t)$. In our case, it reduces to two independent 
 averages: over the amplitudes $x_{i}$ of the spikes and over the waiting times $\tau_{i}$ between them. 
 Noting that $X(t)$ vanishes except if a spike occurs at time $t$, we can express its average 
 in terms of   the average density $S(t)$ of pulses at time $t$ 
 also called the sprinkling time distribution in ref.~\onlinecite{Bardou2002} : 
\begin{equation}\label{eq:average1}
\overline{\langle X(t) \rangle} = h\,S(t)\,.
\end{equation}
Using the expressions (\ref{eq:Sasympt:1<mu<2}), (\ref{eq:Sasympt:mu<1})
for $S(t)$ (see appendix \ref{sec:levy}), we obtain for the average of $X$ :
\begin{subequations}
\begin{align}
\overline{\langle X(t) \rangle}  &= 
\frac{h}{\langle \tau \rangle} = \frac{h}{\tau_{0}} (\mu-1)
&&\quad \text{for} \quad \mu >2,\\
\overline{\langle X(t) \rangle}  &= 
\frac{h}{\langle \tau \rangle} \,\left( 1+\left(\frac{\tau_{0}}{t}\right)^{\mu-1} \right)
&&\quad \text{for} \quad 1<\mu <2,\\
\overline{\langle X(t) \rangle} & =
\frac{\sin (\pi \mu) }{\pi}
\frac{h}{\tau_{0}}
\left(\frac{\tau_{0}}{t} \right)^{1-\mu }
&&\quad \text{for} \quad \mu <1.
\end{align}
\end{subequations}
 Hence the non-stationarity of the noise manifests itself already in the time dependence of this single 
 time average. While it is only a subleading algebraic correction for $1<\mu <2$, 
 this time dependence becomes dominant for $0<\mu<1$. 
  
\paragraph{Two time function}

Following the same lines of reasoning, we can derive the expression of the two time correlation functions 
(with $t>0$) :
\begin{equation}\label{eq:average2}
\langle X(t_p)\,X(t_p+t) \rangle
=h^2\,S(t_p)S(t)\,.
\end{equation}
 The first factor $S(t_{p})$ corresponds to the probability that a spike occurs at time $t_{p}$, while the second
 factor  $S(t)$ reads the probability of having a pulse at time $t_p+t$, knowing that there was one at
$t_p$. This reflects the reinitialization of the noise once a spike has occurred at time $t_p$.
 From (\ref{eq:average1},\ref{eq:average2}), we obtain the connected 
 two points functions: 
 \begin{align}
C(t_{p},t)  &= 
\langle X(t_p)\,X(t_p+t) \rangle_{c} \nonumber \\
&= h^2\, S(t_p)
\left(S(t)-S(t_p+t)\right)\,.
\label{eq:twopoint}
\end{align}

\subsubsection{$1/f$ noise spectrum}

 We will define the spectral density of the noise as the Fourier transform of the connected 
 two points functions restricted to $t>0$ : 
 \begin{equation}
S_X(t_p,\omega)=2\int_0^{+\infty}
C(t_p,t)\,\cos{(\omega t)}\,dt\,.
\end{equation}
 The correlation function $C(t_{p},t)$  generically depends on both times $t_p$ and
$t_p+t$ thus showing that in general $X$ is not a stationary
process. However, to extract the low-frequency behavior of $S_X(t_p,\omega)$, it appears 
sufficient to consider the quasi stationary regime $|t|\ll t_p$ which corresponds 
 to experiment durations much smaller than the age of the noise. In this regime, the connected correlation 
 function (\ref{eq:twopoint}) reduces for $t>0$ to 
 \begin{equation}
\label{eq:def:Ctpt}
C(t_p,t) \simeq h^2  S(t_p)\,\left( S(t)-S(t_p)\right)\,.
\end{equation}
The associated effective power spectrum is defined for 
frequencies $\omega$ large compared to $1/t_p$ and reads:
\begin{equation}
S_X(t_p,\omega) \simeq 
2h^2\,S(t_p)\, \Re{\left(L[S](-i\omega)\right)}
\end{equation}
where $L[S]$ denotes the Laplace transform of $S(t)$.
 Note that in this quasi-stationary regime, the nonstationarity of the noise manifests itself 
 only through the $t_{p}$ dependent amplitude $S(t_{p})$. 
Using explicit expressions for $L[S]$ (see appendix \ref{sec:levy}), 
we obtain the effective power spectra:
\begin{equation}\label{eq:1overf-1}
S_X(t_p,\omega)  \simeq  
\left[h^2\,S(t_p)\,\frac{\cos{(\pi\mu/2)}}{\Gamma(1-\mu)}\right] \,(\omega\tau_{0})^{-\mu}
\end{equation}
 for $0<\mu<1$, and 
\begin{multline}\label{eq:1overf-2}
S_X(t_p,\omega)  \simeq  \\
\left[ h^2\,S(t_p)\,\sin{\left(\frac{\pi(\mu-1)}{2}\right)}(\mu-1)^{\mu-1} \right]
(\omega\langle \tau \rangle)^{\mu-2}
\end{multline}
for the intermediate class $1<\mu<2$.
 The common $1/\omega$ dependence of the spectral density is recovered in the limit $\mu \to 1$. 
 More precisely,  the Laplace transform of $S$ for $\mu=1$, obtained
in appendix \ref{sec:levy}, gives a logarithmic correction to a $1/\omega$ effective power spectrum:
\begin{equation}
S_X(t_p,\omega) \simeq \frac{\pi\,h^2\,S(t_p)}{\omega\tau_{0}\,(\log{(\omega\tau_{0}))^2}}\,.
\end{equation}
Let us stress finally that this effective power spectrum, although useful to compare our approach with 
other noise models, is not sufficient to characterize the statistical properties of the spike field noise
relevant for dephasing. As we shall see in this paper, it 
does not account precisely for the non-stationarity of the 
dephasing factor. Besides this, non Gaussian properties of the noise have strong effects on the dephasing factor
in many regimes. 
 

\begin{figure}
\centerline{\includegraphics[width=8cm]{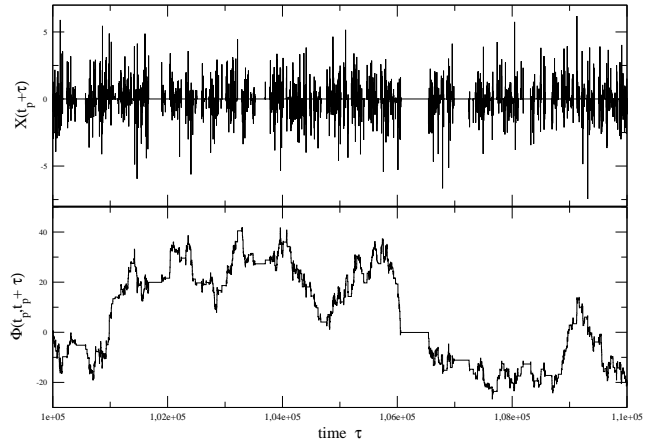}}
\caption{A configuration of the noise $X(t_{p}+\tau)$ in our model (as
a function of $\tau$) between $t_{p}=10^{5}\tau_{0}$
and $t_{p}+\tau=1.1~10^{5}\tau_{0}$. In this
figure, $h=0$ and the waiting times are distributed with an algebraic
distribution $P(\tau_{i})\simeq \tau_{i}^{-2.1}$. The bottom part of
the figure shows the corresponding continuous time random walk of the
accumulated phase of the TLS
between $t_{p}$ and $t_{p}+\tau$.}
\label{fig:noise}
\end{figure}

\section{Dephasing, continuous time random walk of the phase and renewal theory}
\label{sec:RandomWalk}

\subsection{Continuous time random walk of the phase}

Having defined our model for the intermittent noise field $X(t)$, we will now study its effects on the 
dephasing of the two-level system, characterized by the dephasing factor (\ref{eq:defD}). Note that 
{\it a priori} this dephasing has several origins : 
the randomness of the pulses amplitudes $x_{i}$, and the randomness of the waiting times 
$\tau_{i}$. These two dephasing sources are assumed to be independent from each other in this work. 
 For clarity and pedagogical reasons, we first start by considering the effect of randomness of 
 the pulse amplitudes before turning to the the effect of waiting time randomness. 
 
\subsubsection{Dephasing by random phase pulses : the random walk of the phase}

Here, we assume that all waiting times are equal $\tau_{i} = t_{i}-t_{i-1} = \tilde{\tau}$. In this case, the 
phase $\Phi(t_{p},t)$ accumulated between $t_p$ and $t_p+t$ performs a usual random walk
characterized by the distribution $p(x)$ of phase pulses : at the dates $t_{n}=n\tilde{\tau}$, $\Phi(t_{p},t)$ 
is increased by a random value $x_{i}$. 
In the limit $|x|\ll 1$, the phase slowly diffuses
and dephasing is achieved only after a large number of pulses $n=t_{n}/\tilde{\tau}$. Then, the distribution of the 
phase $\Phi(t_{p},t_{n})$ can be well approximated by a Gaussian distribution (apart from some irrelevant tails) and the
dephasing factor is easily computed. It decays exponentially with characteristic time
$\tau_{\phi} = \frac{2\tilde{\tau}}{g^2-h^2}$. 

Note that this diffusive regime can also be studied when the distribution $p(x)$ lacks a finite second 
 or even first moment. An anomalous diffusion of the phase is expected\cite{Bouchaud1990}.
Let us assume that $x_{i}=h+\kappa\xi_{i}$ where the probability distribution of $\xi_{i}$ has zero average
and belongs to the attraction bassin of
the stable law $L_{\nu,\beta}$ characterized by the exponent $0<\nu<2$ of the
 algebraic tails for large values of $\xi_{i}$ and the asymmetry parameter $|\beta| \leq 1$. Then, according to the
 generalized central limit theorem, 
 the accumulated phase after $n$ pulses is  $\Phi_{n}=nh+\kappa\,n^{1/\nu}\xi$ where $\xi$ is distributed according to
 $L_{\nu,\beta}$. Consequently, in the diffusive limit ($n\ll 1$), the dephasing factor is the product of the
 homogenous phase $e^{iht/\tilde\tau}$ by 
 the characteristic function of $L_{\nu,\beta}$ evaluated for 
 $k\simeq g(t/\tilde\tau)^{1/\nu}$. For $\nu<2$ and $\nu\neq 1$, this leads to:
 \begin{equation}
 D_{t_{p}}(t)=e^{-C\kappa^\nu t/\tilde\tau}\,e^{i(h-C\beta \tan{(\pi\nu/2)}\kappa^\nu)\,t/\tilde\tau}
 \,.
 \end{equation}
 where $C$ is a numerical constant which can be absorbed in a rescaling of $\kappa$. 
 Thus, in this case, the decay is still exponential and stationary although non Gaussian features
 of the noise lead to an anomalous dependance of the dephasing time on the coupling constant 
 $\kappa$ that characterizes the scale of the fluctuations of phase pulses\cite{Lutz2002}. 
  
\subsubsection{Dephasing by random waiting times and continuous time random walk}

 Let us now turn to the situation where the phase pulses happen at random times. 
In this case, the accumulated relative phase $\Phi (t_{p},\tau)$ (see eq.\eqref{eq:def-phase}) does not 
perform a random walk  as $\tau$ increases, 
but rather a continuous time random walk\cite{Montroll1965,Haus1987} (CTRW)  on the unit circle. 
In other words, after some random waiting time $\tau_{i}$, $\Phi (t_{p},\tau)$ is incremented by a random value 
$x_{i}$ (see figure \ref{fig:noise}). Thus, on a technical level, the dephasing properties of the two level system 
are now related to some correlation function of the corresponding CTRW. 
 The corresponding dephasing factor can differ from results obtained in the previous section 
 due to the additional source of dephasing given by the randomness of waiting times. 
 To illustrate this point, let us consider the case where all phase pulses have the same intensity $p(x)=\delta(x-h)$.
The accumulated phase after $N$ events is exactly $N h$. Thus, dephasing comes only from the randomness of the number
$N[t_p,t_p+t]$ of events occurring between $t_p$ and $t_p+t$ and the phase diffusion is governed by the probability
distribution for $N[t_p,t_p+t]$. The time needed to obtain $N$ events is the sum over the $N$ first waiting times
after $t_{p}$:
$t_N=\sum_{j=1}^N\tau_j$. At $t_{p}=0$, all $\tau_i$s are distributed according to the same probability distribution
and  therefore, in the limit of large $N$, the generalized central limit theorem\cite{Bouchaud1990} can be used
(the case $t_{p}\neq 0$ will be discussed below).
It provides the limit law for $t_N$ at large $N$ which
in turn determines the probability law for the number of events. 
According to this theorem, three classes must be considered depending on whether the moments 
 \eqref{eq:def-moments} are finite: 
\begin{enumerate}
\item The case $\mu >2$ where both $\langle \tau \rangle$ and $\langle \tau^2 \rangle$ are finite.
In this case, the probability distribution for the number of events is gaussian with a vanishing relative uncertainty.
\item The case $1<\mu < 2$ where the average of $\tau$ is finite but the second moment diverges.
\item The case $\mu<1$ where all moments diverge. 
\end{enumerate}
The usual model for telegraphic noise assumes a Poissonian distribution for the number of events in
a given time interval and corresponds to $\psi(\tau)=\gamma e^{-\gamma\tau}$. We will refer it as the {\it Poissonian case}
and it belongs to the $\mu>2$ class.

 Note that non trivial behavior is expected in the last two cases where $\psi$ has infinite first or second moments. 
In particular, for $\mu <1$, as $\psi(\tau)$ does not have any average, no time scale characterizes the evolution
of the noise and, as we will see, nonstationarity follows. This case deserves a special attention as we showed that 
a $1/f$ spectrum is found precisely for $\mu \to 1$. Before turning to the general formal expressions for the 
dephasing factor, we will focus on the origin of this nonstationarity. 

\subsubsection{Origin of the non-stationarity in CTRW : the first waiting time distribution.}

\begin{figure}[htb]
\begin{picture}(0,0)%
\includegraphics{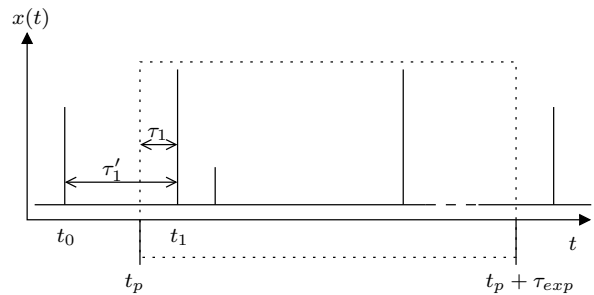}%
\end{picture}%
\setlength{\unitlength}{4144sp}%
\begingroup\makeatletter\ifx\SetFigFont\undefined%
\gdef\SetFigFont#1#2#3#4#5{%
  \reset@font\fontsize{#1}{#2pt}%
  \fontfamily{#3}\fontseries{#4}\fontshape{#5}%
  \selectfont}%
\fi\endgroup%
\begin{picture}(3499,1695)(136,-2527)
\put(811,-2491){\makebox(0,0)[lb]{\smash{{\SetFigFont{8}{9.6}{\familydefault}{\mddefault}{\updefault}{\color[rgb]{0,0,0}$t_{p}$}%
}}}}
\put(2971,-2491){\makebox(0,0)[lb]{\smash{{\SetFigFont{8}{9.6}{\familydefault}{\mddefault}{\updefault}{\color[rgb]{0,0,0}$t_{p}+\tau_{exp}$}%
}}}}
\put(136,-916){\makebox(0,0)[lb]{\smash{{\SetFigFont{8}{9.6}{\familydefault}{\mddefault}{\updefault}{\color[rgb]{0,0,0}$x(t)$}%
}}}}
\put(3466,-2266){\makebox(0,0)[lb]{\smash{{\SetFigFont{8}{9.6}{\familydefault}{\mddefault}{\updefault}{\color[rgb]{0,0,0}$t$}%
}}}}
\put(406,-2221){\makebox(0,0)[lb]{\smash{{\SetFigFont{8}{9.6}{\familydefault}{\mddefault}{\updefault}{\color[rgb]{0,0,0}$t_0$}%
}}}}
\put(1081,-2221){\makebox(0,0)[lb]{\smash{{\SetFigFont{8}{9.6}{\familydefault}{\mddefault}{\updefault}{\color[rgb]{0,0,0}$t_1$}%
}}}}
\put(676,-1816){\makebox(0,0)[lb]{\smash{{\SetFigFont{8}{9.6}{\familydefault}{\mddefault}{\updefault}{\color[rgb]{0,0,0}$\tau'_1$}%
}}}}
\put(946,-1591){\makebox(0,0)[lb]{\smash{{\SetFigFont{8}{9.6}{\familydefault}{\mddefault}{\updefault}{\color[rgb]{0,0,0}$\tau_1$}%
}}}}
\end{picture}%
\caption{\label{fig:waiting-time} Intermittent noise between $t_{p}$ and $t_{p}+t$} 
\end{figure}

Within our model, the waiting times between successive pulses are chosen independently according to the distribution $\psi$. 
Consequently, all the $t_{p}$ dependence of $\Phi(t_{p},t)$ will come from the choice of $\tau_{1}$ defined as the waiting time 
between $t_p$ and the first spike that follows $t_p$ (see figure \ref{fig:waiting-time}).
Indeed, at time $t_{p}+\tau_{1}$, the CTRW starts anew: $\tau_{2}$ is chosen without any correlation to the
history of the CTRW.  Hence given the probability
distribution of $\tau_1$, we can forget about the history of the CTRW of the phase and describe its behavior starting at
$t_{p}$. This remark is at the core of the use of renewal theory. 
In the following, the probability distribution of $\tau_1$ will
be denoted by $\psi_{t_p}$ and
{\it a priori}, it may depend on $t_p$. In fact, as we shall see later, its behavior can be quite counter-intuitive.

First of all, note that a given $\tau_1$ can be obtained from many different noise configurations that differ from the time
of the last event occurring before $t_p$. Separating noise configurations (starting at $t=0$) that have their first spike at time $t_p+\tau_1$
from the others leads to an integral equation that determines $\psi_{t_p}$ in terms of $\psi$ ($S$ is determined from
$\psi$ through an integral equation \eqref{eq:Int-S}):
\begin{equation}\label{eq:integralforpsitp}
 \psi_{t_p}(\tau_1)= \psi(t_{p}+\tau_{1})+ \int_{0}^{t_{p}}d\tau ~ \psi(\tau_{1}+\tau)S(t_{p}-\tau)\,.
\end{equation}
The integral in the r.h.s. comes from noise configurations that have a spike between $0$ and $t_p$. Equation
\eqref{eq:integralforpsitp} is the starting point for deriving analytic results about $\psi_{t_p}$ in appendix
\ref{sec:renewal} and
\ref{sec:explicitpsitp} using Laplace transform techniques. Before computing exactly the dephasing factor, let us
show some of the counter-intuitive properties of $\psi_{t_p}(\tau_{1})$. In particular from 
\eqref{eq:integralforpsitp}, we can derive the following expression for the average of $\tau_{1}$ valid for 
$\mu>2$: 
\begin{equation}
\label{eq:tau1:mu>2} 
\langle \tau_1\rangle _{\psi_{t_p}}=
\frac{\langle \tau^2 \rangle}{2\,\langle \tau \rangle} 
=\frac{\langle \tau \rangle}{2}+
\frac{\langle \tau^2\rangle -\langle \tau \rangle^2}{2\,\langle \tau \rangle}.
\end{equation}
This first term corresponds to the case of regularly spaced pulses averaged over the origin of times. The second term is
the contribution of fluctuations. This result means that irregularities in the event spacings increase the average
waiting time of the first event following $t_p$.
The r.h.s. of \eqref{eq:tau1:mu>2} does not
depend on $t_{p}$, as expected from the stationarity of the CTRW for $\mu>2$ after a short transient regime
at small $t_{p}$.
On the other hand, for $\mu<2$, eq. \eqref{eq:tau1:mu>2} is expected to break down since
$\langle \tau^2 \rangle$ 
diverges. This divergence signals that in some noise configurations 
 $\tau_{1}$ can become of the order of $t_{p}$ and, as a consequence, the average properties of the CTRW after $t_{p}$,
 depend on this age of the noise. Indeed, we can show from \eqref{eq:integralforpsitp} that 
 $\langle \tau_1\rangle _{\psi_{t_p}}$ scales with the age as $t_{p}^{2-\mu}$. 
 Note that in the diffusion regime, many
phase pulses are necessary to dephase the qubit. Therefore, we expect that, in this regime, 
the aging effect on $\psi_{t_p}(\tau_1)$
only brings weak corrections to the dephasing scenario as will be confirmed below by exact computations.
However,  we shall see in the following that the $t_{p}$ dependence of $\psi_{t_p}$ for $\mu\leq1$ has much more
spectacular consequences on the dephasing factor than in the $1<\mu<2$.
 
We will now show that, knowing the Laplace transform of $\psi_{t_p}$, an explicit expression for the Laplace transform of the
average dephasing factor can be obtained.
 
\subsection{Exact dephasing via renewal theory}
\label{sec:Exact}

\subsubsection{Dephasing factor}

Among all noise configurations that are to be taken into account, some of them (possibly very few) 
do not have any event between $t_p$ 
and $t_p+t$. Their total weight is given by $\Pi_0(t_p,t)$ which is
\begin{equation}
\Pi_0(t_p,t) = \int_t^{+\infty}\psi_{t_p}(\tau)\,d\tau.
\end{equation}
All other histories have at least one event between $t_p$ and $t_p+t$. Let
us assume that it happens at $t_p+\tau$ where $\tau$ lies between
$0$ and $t$. Then after this event, the noise starts anew. The jump itself
contributes by $\overline{e^{ix}}$ to the dephasing factor and the rest of the noise configuration contributes
by $D_0(t-\tau)$ ({\it i.e.} $t_{p}=0$ in eq.(\ref{eq:defD})).  
The probability that the first event after $t_p$ happens
at time $t_{p}+\tau$ is nothing but $\psi_{t_p}(\tau)$. Hence the contributions to $D_{t_{p}}(t)$ 
from all possible noise configurations take the form of the following renewal equation: 
\begin{equation}
\label{eq:integralDtp}
D_{t_{p}}(t) = \Pi_0(t_p,t) +\overline{e^{ix}}\,\int_0^t
\psi_{t_p}(\tau)D_0(t-\tau)\,d\tau.
\end{equation}
Note that the Laplace transform of this expression is very simple:
\begin{equation}
\label{eq:DysonD}
L[D_{t_{p}}] = L[\Pi_0] +(1-f )\left(L[\psi_{t_p}]\ldotp L[D_0]\right).
\end{equation}
where $f =1-\overline{e^{ix}}$. Specializing $t_p=0$, one gets an expression for $D_0(t)$ that
contains $\Pi_0(0,t)$. Since $\Pi_0(0,t)=\int_t^\infty \psi(\tau)\,d\tau$ an explicit expression for the Laplace transform
of $D_0$ can be found:
\begin{equation}
\label{eq:LD0}
L[D_0]=\frac{1}{s}\, \frac{1-L[\psi]}{1-(1-f )\,L[\psi]}.
\end{equation}
Plugging expression \eqref{eq:LD0} into equation \eqref{eq:DysonD} gives:
\begin{equation}
\label{eq:resultat:LD}
L[D_{t_p}] =
\frac{1}{s}\left(1 - \frac{f \,L[\psi_{t_p}]}{1-(1-f )L[\psi]}
\right)\,.
\end{equation}
This exact expression will be extensively used to derive both 
analytic expressions and numerical plots by Laplace inversion. 
Before proceeding along, let us express $L[D_{t_p}]$ in terms of
the density of events. It is then convenient to define
$S_{t_p}(t)=S(t_p+t)$ (see appendix \ref{sec:levy}). From the expressions in this appendix of 
$L[S_{t_p}]$ and $L[S]$  in terms of $L[\psi]$ and $L[\psi_{t_p}]$, we obtain 
\begin{equation}
\label{eq:resultat:LD2}
L[D_{t_p}] =
\frac{1}{s}\left(1-\frac{f \,L[S_{t_p}]}{1+f \,L[S]}\right).
\end{equation}
Equation \eqref{eq:resultat:LD} and \eqref{eq:resultat:LD2}
enable analytical estimates of the dephasing factor in various limiting
regimes. Equation \eqref{eq:resultat:LD2} is useful in the limit of small
coupling ({\it i.e.} $f \rightarrow 0$) whereas eq. \eqref{eq:resultat:LD} will
be useful in the opposite limit of a wide distribution of spikes' heights
({\it i.e.} $f \rightarrow 1$).

\subsubsection{Weak and strong coupling regimes}
\label{sec:WeakStrong}

In the limit of very strong coupling ($f =1$), the phase
spreading of a single spike is sufficient to dephase the qubit. In
this case, eq.\eqref{eq:integralDtp} immediately
leads to $D_{t_p}(t)=\Pi_0(t_p,t)$. This means that
the whole average is dominated by the rare noise configurations that
do not evolve during the experiment. As we shall see later, this
leads to a strong $t_p$ dependence of $D_{t_p}(t)$ for $\mu<2$.

At much lower couplings ($|f |\ll 1$),
it is not a priori clear whether the renewal
equation for $D_{t_p}(t)$ induces a strong dependence of
$D_{t_p}(t)$ on $t_p$. Pushing the above analysis forward
amounts to compare the time dependence of the 
two terms in the r.h.s. of the renewal equation \eqref{eq:integralDtp}. As we will see now,
it will provide a better understanding of the physics underlying the dephasing
scenario.

Let us first assume that $\Pi_0(t_p,t)$ decays much faster than
$D_0(t)$. Since $\psi_{t_p}(\tau)=-\partial_\tau\Pi_0(t_p,\tau)$, it also decays much faster 
than $D_{0}$. Consequently, we can approximate 
 $\psi_{t_p}(\tau)D_{0}(t-\tau)$ by $\psi_{t_p}(\tau) D_{0}(t)$ in Eq.(\ref{eq:integralDtp}). 
Then, after a
short initial decay due to both $D_{0}(t)$ and $\Pi_0(t_p,t)$,
$D_{t_p}(t)$ decays as $D_0(t)$. Consequently, the dephasing time
does not depend on $t_p$.  For instance, we expect this situation to occur at
weak coupling when the average waiting time $\langle \tau \rangle$ is finite
since the probability that no event occurs between $t_p$ and
$t_p+t$ vanishes quite fast when $t$ increases.
Note that in the limit of vanishing coupling, the spreading of the phase can become
much slower than the decay of $\Pi_0(t_p,t)$ and thus our
starting point hypothesis is valid.

The opposite case where the decay of $D_0(t)$ is much faster than
the decay of $\Pi_0(t_p,t)$ can be discussed more conveniently by
integrating the renewal equation \eqref{eq:integralDtp}
by parts. Considering again $g\ll 1$ we get:
\begin{equation}
\label{eq:DtpD0Pi0}
D_{t_p}(t) \simeq D_0(t) - \int_0^t
\Pi_0(t_p,t-\tau) D'_0(\tau)\,d\tau\,.
\end{equation}
Approximating $\Pi_0(t_p,t-\tau)\approx \Pi_0(t_p,t)$
in this equation yields $D_{t_p}(t) = D_0(t) + \Pi_0(t_p,t)
(1-D_0(t))$. As a consequence, once $D_{0}(t)$ has decayed, the dephasing factor is almost equal
to $\Pi_0(t_p,t)$. This is the same behavior than in the very strong coupling regime $f 
=1$, although here, we assumed $|f | \ll 1$. This regime is expected to occur
when the decaying time scales of $D_0(t)$ and of
$\Pi_0(t_p,t)$ are comparable. This is obviously the case at
very strong coupling but, surprisingly, as we shall see now it can also be obtained for $|f |\ll 1$! 

First of all, the
above discussion shows that such a regime can only happen if the average waiting time is
infinite, {\it i.e.} for $\mu<1$.
In this case, $\Pi_0(t_p,t)$ can be evaluated analytically (see eq. \eqref{eq:Pi0:0<mu<1}):
it is shown to be independent of $\tau_{0}$ and to exhibit aging behavior
({\it i.e.} to depend only on $t/t_p$). Therefore, the decaying time scale of $\Pi_0(t_p,t)$
scales as $t_p$. Comparing this time scale with the
dephasing time scale for $|f |\ll 1$, leads to a 
$t_p$ dependent cross-over coupling constant $f_c(t_p)$. In the case of aging $\Pi_{0}$,
the crossover coupling decays to lower and lower values by increasing $t_{p}$.

For $|f|\gtrsim f_c(t_{p})$,  the dephasing
factor behaves as $\Pi_0(t_{p},t)$ and the dephasing time saturates as a function of $f$. 
Such a saturation of the dephasing time as a function of the
amplitude of the noise has already been discussed for a
Poissonian fluctuator\cite{Falci2003,Galperin2003}.
In this case
$\Pi_0(t_p,t)=\Pi_0(0,t) = e^{-t/\langle \tau \rangle}$ decays very fast, on a time scale
$\langle \tau \rangle$. The cross-over between weak and strong coupling regime
happens precisely at the point where the dephasing time assuming
weak coupling $\langle \tau \rangle/g^2$ is of the same order as this decay time,
i.e. for $g\sim 1$. Note that in this case, as expected from our
discussion, the crossover scale is independent of the age of the
noise $t_p$. We expect this reasoning to break down in our model
because of the broad distribution of waiting times ($\mu<2)$. Understanding the
various dephasing scenarii and computing the $t_p$ dependence of
the crossover coupling in our case requires the computation of
$\Pi_0(t_p,t)$ and of $D_0(t)$. These quantitative results will be presented in
forthcoming sections.

To summarize the above discussion, we have argued that the dephasing time
is bounded by the typical decay time of both $\Pi_0(t_p,t)$ and $D_0(t)$.
This suggests to distinguish between two regimes:
on one hand, a {\em weak coupling regime} for which $D_0(t)$ decays much slower than
$\Pi_0(t_p,t)$ and for which the dephasing time - in that case just the decay time
of $D_0(t)$ - is not sensitive to $t_p$. On the other hand, a {\em strong coupling regime}
for which $D_0(t)$ decays faster than $\Pi_0(t_p,t)$. In that case, the
dephasing time is given by the decay time of $\Pi_0(t_p,t)$ and thus possibly
$t_p$-dependent. As the above discussion shows, the cross-over between these
two regimes may happen for a possibly $t_p$ dependent cross-over
coupling $f_c(t_p) \ll 1$. Note
that in the strong coupling regime, the dephasing time becomes independent of the
amplitude of the noise whereas in the weak coupling regime, it is expected to depend
on the amplitude and to diverge in the vanishing coupling limit.

%
\section{Dephasing in the symmetric models}
\label{sec:symmetric}

In this section we present our results for the situation of a
symmetric distribution of the spikes, $p(-x)=p(x)$. 
Under this assumption the average random accumulated phase vanishes, {\it i.e.}
$f $ and consequently $D_{t_p}(t)$ are real. For $g\ll 1$,
$f $ may be expanded in moments of $p(x)$, $f  \approx g^2 \ll 1$
where $g$ measures the typical scale of the fluctuations of the spikes.

\subsection{Dephasing at $t_p=0$}
\label{sec:symmetric:mu>2}

Before discussing the decoherence factor for arbitrary preparation time
$t_p$, it is useful to investigate the case $t_p=0$.
Rewriting the Laplace transform of $D_0(t)$ \eqref{eq:LD0} as
\begin{equation}
\label{eq:LD0b}
L[D_0](s) = \frac{1}{s}\left(1+\frac{f  L[\psi]}{1-L[\psi]}\right)^{-1}
\end{equation}
suggests to introduce a scale $\gamma_g$ related to the strength of the coupling:
\begin{equation}
  \label{eq:tg}
  \left.\frac{f  L[\psi]}{1-L[\psi]}\right|_{s=\gamma_{g}} = 1\,.
\end{equation}
Note that $\gamma_g$ vanishes with the coupling $g$.  Investigating the
behavior of $D_0$ for $t\ll \tau_{0}$ requires evaluating the Laplace
transform for $s\tau_{0}\ll 1$. In this regime,
the Laplace transform $L[\psi]/(1-L[\psi])$ can be approximated
by $1/(1-L[\psi])$. Within this approximation, we shall now
derive explicit expressions for $L[D_{0}]$ which can be Laplace inverted
explicitly. This leads to expressions for
$D_0(t)$ valid at $t\gg \tau_{0}$ for the different classes of $\mu$.

In the case of finite average waiting time ($\mu>1$), we can expand
$L[\psi] \approx 1 - s\langle \tau \rangle$ to find the leading contribution
to $D_0(t)$ for $\langle \tau \rangle\ll \gamma_{g}t< 1$. This gives:
\begin{equation}
  \label{eq:D0mu>1}
  D_{0}(t) \simeq
L^{-1}\left[\frac{\langle \tau \rangle}{s\langle \tau \rangle+f }\right]
  = e^{-t/\tau_{\phi}} \;,
\end{equation}
with the dephasing time $\tau_\phi = \gamma_{g}^{-1}=\langle \tau \rangle/f $ (see eq. \eqref{eq:tg}).
Note that this expression is exact only for 
$\psi(\tau) = \frac{1}{\langle \tau \rangle} e^{-\tau/\langle \tau \rangle}$.
At finite non integer $\mu$, taking into account the fluctuations of the waiting times
requires keeping all terms in $1-L[\psi]$ up to the first non integer power
$(s\tau_{b})^\mu$.
For $1<\mu<2$, we get algebraic subleading corrections to
\eqref{eq:D0mu>1}:
$\log{(D_{0}(t))} \simeq -\gamma_{g} t  - f  c(\mu) (t/\tau_{0})^{2-\mu}$, where
$c(\mu) = (\mu-1)/(2-\mu)$. These corrections being weak for $g\ll 1$, the dephasing time 
$\tau_{\phi}$ remains equal to $\gamma_{g}^{-1}=\langle \tau \rangle/g^2$ in this regime.

For $\mu < 1$, the first term in the expansion of $1-L[\psi]$
is proportional to $(\tau_{0} s)^\mu$. Therefore,
plugging in $(1-L[\psi])/L[\psi] \simeq \Gamma(1-\mu) (s\tau_{0})^\mu$
and performing the inverse Laplace transform
of \eqref{eq:LD0b} gives:
\begin{equation}
  \label{eq:D0mu<1}
  D_0(t) \simeq
L^{-1}\left[\frac{s^{-1}}{1+f (s\tau_{0})^{-\mu}}\right]
  = E_\mu\left[-\left(\gamma_{g} t \right)^\mu\right]
\end{equation}
where $E_\mu(z) = \sum_{n=0}^\infty\frac{z^n}{\Gamma(1+\mu n)}$ denotes
the Mittag-Leffler function\cite{Bateman1955}. For values of $\mu$ close to one and
$z\lesssim 1$ this function can be approximated by a simple exponential,
whereas for $\mu \sim 0$ it is similar to an algebraic function
$E_\mu(z)\sim (1+z)^{-1}$. For large values of the argument
($\abs{z} \gg 1$) and $\abs{\mathrm{arg}(-z)}<(1-\mu/2)\pi$,
we obtain\cite{Bateman1955} $E_\mu(z)\approx (-z \Gamma(1-\mu))^{-1}$. 
This change of behavior from an exponential to an algebraic behavior was 
interpreted in ref. \onlinecite{Schriefl2004} as a Griffith effect. 

Computing long time behavior ($\gamma_{g} t \gg 1$) of the dephasing factor
can be done by expanding \eqref{eq:LD0b} as follows:
\begin{equation}
  \label{eq:LD0lt}
  L[D_0](s) \simeq \frac{1}{s}\frac{1-L[\psi]}{f  L[\psi]}\;.
\end{equation}
For $\gamma_{g} t\gg 1$  ($s  \ll \gamma_{g}$) we can safely replace
$L[\psi] \approx 1$ in the denominator. The inverse Laplace transform can
then be done easily and reads for $\gamma_{g}t\gg 1$ :
\begin{equation}
\label{eq:D0lt}
D_0(t) \simeq \frac{1}{f }\int_t^\infty \psi(\tau)\,d\tau  =
\frac{1}{f }\left(\frac{\tau_{0}}{\tau_{0}+t}\right)^\mu \;.
\end{equation}
For $\mu<1$, \eqref{eq:D0lt} is nothing but the asymptotic
behavior of \eqref{eq:D0mu<1} for $\gamma_{g}t\gg 1$.
Note that for $g\gg 1$, $\gamma_g^{-1}$ is of the
order of $\tau_{0}$ and the decay is algebraic at all times $t>\tau_{0}$,
given by \eqref{eq:D0lt}. For $1<\mu<2$ and $g\ll 1$, only at large times
($\gamma_{g}t \sim \ln(1/f )$), when the qubit has almost completely dephased,
the decay crosses over to the power law \eqref{eq:D0lt}.

\subsection{Influence of a finite preparation time $t_p$}
\label{sec:Dtp}

We will now discuss in detail the $t_p$-dependence of the dephasing
scenario and of the crossover coupling strength $g_c$ for the different classes of $\mu$.
Simple analytical expressions for $D_{t_p}(t)$ can be derived
in the weak $(g\ll g_c)$ and strong $(g \gg g_c)$
coupling regimes.

\subsubsection{Infinite fluctuations of the waiting times: $1<\mu<2$}
\label{sec:1<mu<2}

\begin{figure}
\includegraphics[width=8cm]{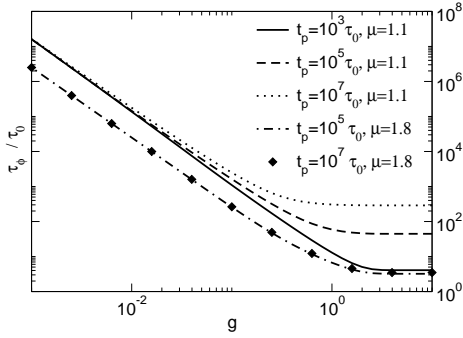}
\caption{\label{fig:tauphi:1<mu<2} Dephasing time $\tau_\phi$ in the symmetric model as a
function of $g$ for $\mu=1.1$, $\mu=1.8$ and for various values of $t_p$. At
strong coupling the dephasing time exhibits
$t_{\mathrm{p}}$-dependence, which disappears as $\mu$ increases
to higher values. Note that the critical coupling $g_c$ has a
weak $t_{\mathrm{p}}$-dependence for $\mu$ close to one.}
\end{figure}

For $1<\mu<2$, the decay of $D_0(t)$ is accurately described by \eqref{eq:D0mu>1} and
\eqref{eq:D0lt}.
On the other hand, $\Pi_0(t_p,t)$ exhibits an explicit dependence on
the age of the noise (see appendix \ref{sec:Pi0}):
\begin{equation}
  \label{eq:Pi01<mu<2}
  \Pi_0(t_p,t) \simeq \left(\frac{\tau_{0}}{\tau_{0}+t}\right)^{\mu-1} -
   \left(\frac{\tau_{0}}{\tau_{0}+t+t_p}\right)^{\mu-1}
\end{equation}
for $t_p\gg \tau_{0}$. 
 We will arbitrarily define the typical time scale of the decay of any function as 
 the time at which its modulus reaches a fixed value $0<\alpha<1$ ($\alpha=1/e$ in
all the figures of this paper). 
The cross-over between a weak and strong coupling regime is defined as
the value of $g$ where the typical
decay times of $D_0(t)$ and $\Pi_0(t_p,t)$ coincide (see section \ref{sec:WeakStrong}).
In the present case, the cross-over coupling gets a weak $t_p$ dependence:
\begin{equation}
  \label{eq:gc1<mu<2}
  (g_c(t_p))^2 \simeq \frac{1}{\mu-1}\left[\alpha +
    \left(\frac{\tau_{0}}{t_p}\right)^{\mu-1} \right]^{1/(\mu-1)}  \;.
\end{equation}
Note that $g_c$ is a decreasing function of $t_p$ since increasing $t_p$
slows down the decay of $\Pi_0(t_p,t)$ (remember that
the average time of the first occurring spike after $t_p$
increases as $t_p^{2-\mu}$). But since the
average waiting time is finite, $g_c$ has a non zero lower bound.
Note also that the $t_p$-dependence of $g_c$ is only visible
for values of $\mu$ close to one and disappears as $\mu$
increases to higher values. This can be seen on the numerical
results depicted on fig.~\ref{fig:tauphi:1<mu<2}: $g_c(t_p)$ is
the crossover coupling where the dephasing time start to saturate
as a function of $g$.

\begin{figure}
\includegraphics[width=8cm]{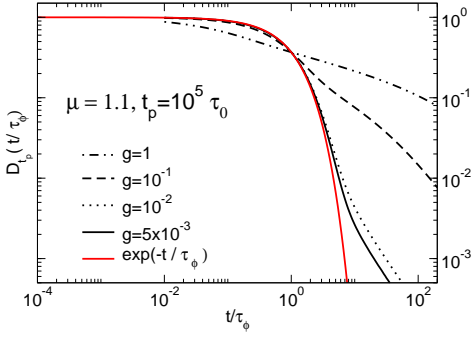}
\caption{\label{fig:Dtp:1<mu<2} Dephasing factor $D_{t_p}(t)$ in the symmetric model
obtained by numerical Laplace inversion for
$\mu=1.1$ and $t_p=10^5$ and various values of $g$. At small
couplings, the decay is exponential. But at longer times
$t\gg \tau_\phi$ and for strong coupling the decay is algebraic.}
\end{figure}

In the case of very strong coupling ($g\gg g_c(t_p)$), the decay of $D_{t_p}(t)$
coincides with $\Pi_0(t_p,t)$ and is thus algebraic at all times
(see fig. \ref{fig:Dtp:1<mu<2}).
On the other hand, at weak coupling ($g \ll g_c$),
the dephasing factor $D_{t_p}(t)$ follows
$D_0(t)$ up to times of the order of the dephasing time,
{\it i.e.} it decays exponentially:
$D_{t_p}(t) = \exp(-t/\tau_\phi)$, where $\tau_\phi= \gamma_{g}^{-1} = \tau_{0} g^{-2}$.
In this weak coupling regime, the dephasing
is thus described accurately using a second cumulant expansion.
However, for times large compared
to the dephasing time, $t \gtrsim \tau_\phi \ln(\tau_\phi/\tau_{0})$,
higher cumulants contribute and the decay goes over to a much slower
power law. If $t_p < \tau_\phi$ the asymptotic decay of $D_{t_p}(t)$
continues to follow $D_{0}(t)$ behavior given in \eqref{eq:D0lt} whereas, 
in the opposite case $t_p>\tau_\phi$
it is given by $\Pi_0(t_p,t)$ \eqref{eq:Pi01<mu<2}.

\subsubsection{Infinite average waiting time:  $0<\mu<1$}
\label{sec:mu<1}

\begin{figure}
\includegraphics[width=8cm]{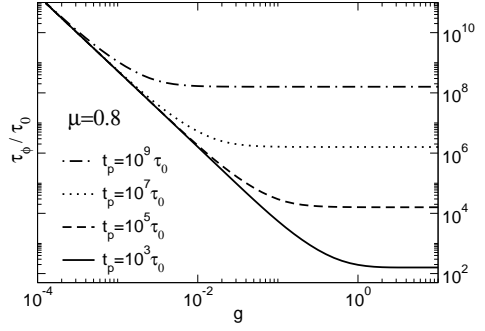}
\caption{\label{fig:tauphi:mu<1} Dephasing time $\tau_\phi$ in the symmetric model as a
function of $g$ for $\mu < 1$ and for various values of $t_p$. At
strong coupling the dephasing time exhibits a
$t_{\mathrm{p}}$-dependence, which disappears as $\mu$ increases
to higher values. Note the explicit $t_{\mathrm{p}}$-dependence of
the critical coupling $g_c$.}
\end{figure}

For $0<\mu<1$ the influence of a finite preparation time
$t_p$ becomes even more drastic, due to the absence of a characteristic
time scale in the waiting time distribution.
In this case $\Pi_0(t_p,t)$  only depends on the
ratio $t/t_p$ (aging behavior):
\begin{equation}
\label{eq:Pi0mu<1} \Pi_0(t_p,t) =
\frac{\sin{(\pi\mu)}}{\pi\mu}\,\left(
\frac{t_p}{t}\right)^\mu\,{}_2 F_1(1,\mu;1+\mu;-\frac{t_p}{t}) \;,
\end{equation}
where ${}_2F_1$ denotes a hypergeometric function\cite{Gradsteyn1980}. Consequently, the
typical decay time of $\Pi_0(t_p,t)$ is proportional to $t_p$.
On the other hand, $D_0(t)$ is given by \eqref{eq:D0mu<1} and its
decay time thus scales as $\tau_{\phi}=\gamma_{g}^{-1} \propto \tau_{0}/f ^{1/\mu} \approx \tau_{0}/g^{2/\mu}$
in the limit $g\ll 1$.
As a consequence, the crossover coupling strength $g_c$ exhibits a strong
$t_p$ dependence:
\begin{equation}
  \label{eq:gcmu<1}
  g_c(t_p) = \lambda(\mu) \left(\frac{\tau_{0}}{t_p}\right)^{\mu/2} \ll 1 \,,
\end{equation}
where $\lambda(\mu)$ is a function of $\mu$ that can be obtained by inversion of
\eqref{eq:Pi0mu<1} and \eqref{eq:D0mu<1}.
As in the case $1<\mu<2$, $g_c$ is a decreasing function of $t_p$: the
range of the strong coupling regime increases with the age of the noise.
However, contrarily to the case $1<\mu<2$, $g_c$ has no lower bound, {\it i.e.} it
decays to zero as we increase $t_p$. Consequently, any qubit surrounded
by a noise with $0<\mu<1$ will eventually end up in the strong coupling regime.
This can be seen on results depicted on figure \ref{fig:tauphi:mu<1}.

\begin{figure}
\includegraphics[width=8cm]{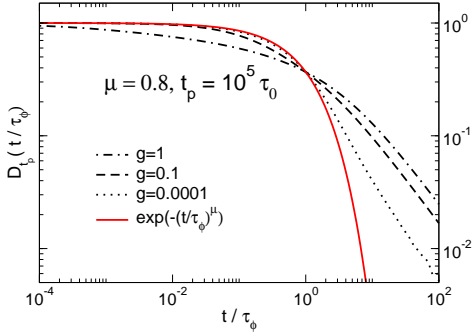}
\caption{\label{fig:Dtp:mu<1} Dephasing factor $D_{t_p}(t)$ in the symmetric model
obtained by numerical Laplace inversion for
$\mu=0.8$, $t_p=10^5$ and various values of $g$. For weak
coupling, $g<g_c(t_{\mathrm{p}})$, the decay is exponential for
$t<\tau_\phi$. For times $t\gtrsim \tau_\phi$ at weak coupling
and for strong coupling the decay is algebraic.}
\end{figure}

At strong coupling ($g>g_c(t_p)$), $D_{t_p}(t)$ is given by $\Pi_0(t_p,t)$
and therefore is only a function of $t/t_p$. In this
regime, the dephasing time is proportional to $t_p$, as shown
on fig. \ref{fig:Dtp:mu<1}.
The initial decay of coherence is quite fast since, for $t\ll t_p$,
$\Pi_0(t_p,t) \approx 1 - A(\mu) (t/t_p)^{1-\mu}$
with $A(\mu)=\sin(\pi\mu)/[(1-\mu)\pi]$. Consequently, for $\mu$ close
to one $D_{t_p}(t)$ decays substantially for times short
compared to the preparation time $t_p$. For
$t\gtrsim t_p$, the decay slows down considerably and goes over to a power law
$\Pi_0(t_p,t) \propto (t_p/t)^\mu$.

At weak coupling ($g<g_c(t_p)$), the decay of $D_{t_p}(t)$
is accurately described by \eqref{eq:D0mu<1}.
For $\mu$ close to 1 and $\gamma_{g}t< 1$, the Mittag-Leffler function
can be approximated by a  exponential
decay $D_{t_p}(t) \approx \exp[-(\gamma_{g}t)^\mu]$, whereas for
$0<\mu \ll 1$ and $\gamma_{g}t\lesssim 1$, the decay is rather algebraic,
$D_{t_p}(t) \approx [1+(\gamma_{g}t)^\mu]^{-1}$.
In any case, the typical decay time scales as
$\tau_\phi =\gamma_{g}^{-1}  \propto g^{-2/\mu}$.
As for $1<\mu<2$, this dephasing time
can be recovered using a second cumulant expansion of the
phase. Obviously, for larger times, $t>\tau_\phi$, higher cumulants contribute
and the decay goes over to a power law,
$D_{t_p}(t) \propto (\tau_{0}/t)^{\mu}$, as shown in Fig \eqref{fig:Dtp:mu<1}.

\subsubsection{On the marginal case: $\mu=1$}

Finally, in the marginal case $\mu=1$ corresponding to $1/f$ noise, the
above analysis is confirmed qualitatively. We mainly find logarithmic
corrections to the above results, leading to behaviors intermediate between the classes
$\mu<1$ and $1<\mu<2$. The crossover coupling $g_c$ in this case turns out to be
\begin{equation}
  \label{eq:gcmu=1}
  g_c(t_p) = \left(\frac{\tau_{0}}{t_p} \ln(t_p/\tau_0)\right)^{1/2}\,,
\end{equation}
showing a strong dependence on the preparation time $t_p$. As in the case
$\mu<1$, it decays to zero for $t_p\to \infty$.

Computing $D_0(t)$ requires to invert \eqref{eq:LD0b}. This
cannot be done exactly because of the complicated form of
$L[\psi] \simeq 1-s\tau_{0} \abs{\ln(s\tau_{0})}$ for $s\tau_{0} \ll 1$.
However, $D_0(t)$ may be estimated considering $\ln{(s\tau_{0})}$ as a constant and 
replacing  $s$ by $1/t$ in its argument. This leads to:
\begin{equation}
  \label{eq:D0mu=1}
D_0(t) \simeq \exp\left(-f  \frac{t/\tau_{0}}{\ln(t/\tau_{0})}\right) \quad 
\mathrm{for}\ 
t\lesssim \frac{\tau_{0}}{f }\abs{\ln f }  \;.
\end{equation}
Hence, for weak coupling, $D_{t_p}(t)$ decays due to the contribution of
$D_0(t)$ on a characteristic time scale $\tau_\phi \simeq (\tau_{0}/f ) \abs{\ln f }$.
Again, for times large compared to the dephasing time,
the decay slows down considerably and goes over to a power law \eqref{eq:D0lt},
$D_{t_p}(t) \propto t^{-1}$.

At strong coupling ($g > g_c(t_p)$), $D_{t_p}(t)$ follows $\Pi_0(t_p,t)$, which
reads in the present case:
\begin{equation}
  \label{eq:Pi0mu=1}
  \Pi_0(t_p,t) \simeq \frac{1}{\ln(1+t_p/\tau_{0})}
               \ln\left(1+\frac{t_p}{\tau_{0}+t}\right) \;.
\end{equation}
Consequently, the dephasing time exhibits a strong dependence on the preparation
time $t_p$. However, since $\Pi_0(t_p, t)$ is not a function of $t/t_p$
the dephasing time does not scale linearly with $t_p$ (as for $\mu<1$) but
has a weaker $t_p$ dependence which depends on the parameter $\alpha$ used to define $\tau_\phi$ 
($D_{t_p}(\tau_\phi) = \alpha$).

\section{Dephasing in the asymmetric models}
\label{sec:Asymmetric}

In this section, we consider asymmetric distributions $p(x)$ of the pulse amplitudes 
with a non vanishing average $h=\overline{x}$. This obviously 
corresponds to the generic case for our intermittent noise, but it also appears naturally in the context 
of decoherence at optimal points discussed in section \ref{sec:decoherence}. 

Similar expansions of the exact expression \eqref{eq:resultat:LD}
can be performed to derive the detailed decay of $D_{t_p}(t)$ in the case of a finite
mean value of the noise, $\overline{x}\neq 0$.
Nevertheless, analytic expressions are much harder to obtain since
in this case $f =1-\overline{e^{ix}}$ is complex. Therefore, in this section, we will mainly derive scaling
laws of $\tau_\phi$, the critical coupling $g_c$ and present numerical computations of $D_{t_p}(t)$ in 
various regimes. 
 Note that at strong coupling, the dephasing is due to single events of the noise,
{\it i.e.} $D_{t_p}(t) \simeq \Pi_0(t_p,t)$.
As a consequence, dephasing only depends on $\psi(\tau)$, not on the details
of $p(x)$. In this regime, the results reduce to those previously derived for the symmetric case
($h=\overline{x} = 0$) in the strong
coupling regime. Differences with respect to the symmetric case  only arise
in the weak coupling regime on which we shall focus in the following.

 Before proceeding along, let us  recall the notations \eqref{eq:moments-p}  for the moments of $p(x)$ : 
$h=\overline{x}$ and $g^2=\overline{x^2}$.  
At weak coupling, $f $ may be expanded in moments
of $p(x)$: $f  \approx - ih + g^{2}/2 + \dots$.
As expected for  $h \ll g^{2}$, the finite
mean value does not modify the results of the previous section, {\it i.e.}
the dominant dephasing is the same as in the $h=0$ case (symmetric noise). However, as 
$h$ gets of the same order as $g^{2}$
the decay of $D_{t_p}(t)$ and the scaling of $\tau_\phi$ are
modified. Understanding precisely the possibly nonlinear effect of a small bias
on the dephasing factor is related to the fluctuation/dissipation issue in
CTRW and is out of the scope of the present paper.
Therefore we shall now focus on the case of huge asymmetry
where $h \gg g^{2}$. In this limiting regime, the pulse dispersion
can be forgotten and $h$ is the only coupling parameter. Note that in this specific
variant of the asymmetric model, 
$\Phi_{t_{p}}(t)$ is proportional to the number of events $N[t_{p},t_{p}+t]$ that occur between
$t_{p}$ and $t_{p}+t$ and therefore, the dephasing factor as a function of $h$ is the characteristic
function of the probability distribution for $N[t_{p},t_{p}+t]$. 
Contrarily to the case of Poissonian telegraph noise\cite{Paladino2002,Galperin2003} where a finite
mean value of the noise just adds a global phase to $D_{t_p}(t)$, we will see that strong 
fluctuations in the occurrence times of the spikes induce dephasing even for a fixed value of the phase pulses.

\subsection{The $0<\mu<1$ class}

\begin{figure}[ht]
  \centering
  \vspace{.5cm}
 \includegraphics[width=8cm]{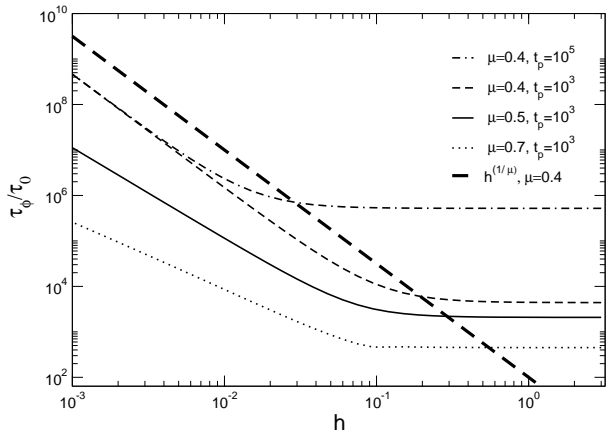}
  \caption{
Dephasing time $\tau_\phi$ in the asymmetric model as a function of the coupling $g$ for
a huge asymmetry $h \gg g^{2}$, and different values
of $\mu<1$. At weak coupling, $h<h_c(t_p)$,
the dephasing time scales as $\tau_\phi \propto 1/h^{1/\mu}$, whereas
it becomes proportional to $t_p$ for $h>h_c(t_p)$.
Note the $t_p$-dependence of the critical coupling $h_c$, visualized for
$\mu=0.4$.
}
  \label{fig:asymtaumu<1}
\end{figure}

In the $0<\mu<1$ case, we can use the same method as for the derivation of
\eqref{eq:D0mu<1} to find the scaling law of $\tau_\phi$ as a function of
the coupling strength.
The dephasing factor in the weak coupling regime decays as
$\abs{D_{t_p}(t)}=\abs{E_\mu(i z)}$ where
$z \simeq -h (t/\tau_{0})^\mu/\Gamma(1-\mu)$
and
$E_\mu$ denotes the Mittag-Leffler function previously used.
Using the series expansion that defines $E_\mu$, it can be shown easily that
$\abs{E_\mu(i z)}$ only depends on $z^2$. As a consequence,
the dephasing time in the weak coupling regime scales as
$\tau_\phi \propto \tau_{0}/h^{1/\mu}$, as shown in
Fig.~\ref{fig:asymtaumu<1}.

The cross-over between weak and strong coupling regime is obtained by 
comparing this time scale with the typical decay time of $\Pi_0(t_p,t)$,
which is proportional to $t_p$ for $0<\mu<1$.
Therefore the critical coupling strength scales as
$h_c \propto t_p^{-\mu}$.

\begin{figure}[ht]
  \centering
   \vspace{.5cm}
  \includegraphics[width=8cm]{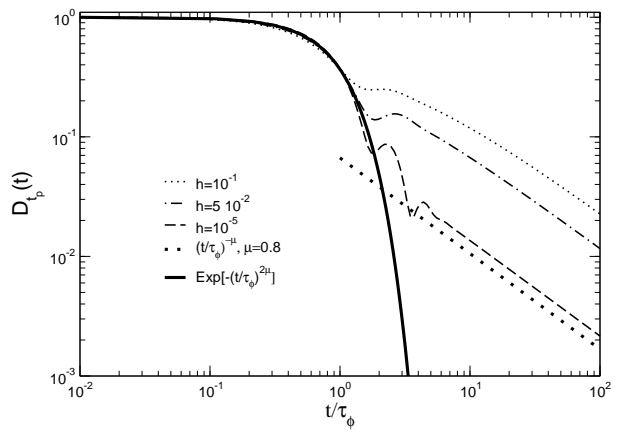}
  \caption{
  Decay of $\abs{D_{t_p}(t)}$ in the asymmetric model 
  for $\mu=0.8$, $t_p=10^3\tau_{0}$ and different values of the
coupling strength $h$.
  Here $h_c(10^3\tau_{0})\simeq 10^{-2.4}$}
  \label{fig:qsdqsd}
\end{figure}

Fig. \ref{fig:qsdqsd} shows $D_{t_p}(t)$ for $\mu=0.8$ and
different values of the coupling
strength from the weak coupling regime to slightly above
$h_c$. The initial decay, up to the dephasing time,
follows the previous exponential decay. At larger times
the decay crosses over to a power law, $D_{t_p}(t) \propto 1/t^\mu$,
as in the symmetric case.
Between these two asymptotic behaviors, beatings may occur, due to
interferences between different noise histories, as already seen in other models \cite{Falci2003}. However,
in this reference, oscillations appear in the strong coupling regime. They arise from interferences between 
noise histories associated with different initial conditions. Such a strong initial condition dependence is 
expected at strong coupling for a Poissonian fluctuator. In our case, oscillations are also present at $t_{p}=0$ 
in the diffusion regime and
therefore cannot be explained by an initial condition dependence. Nevertheless they are still associated with an
interference effect between noise histories.
 
To illustrate this point, let us mention that for $D_{0}(t)$, a numerical investigation shows that in the situation considered 
here ($h\gg g^{2}$), these oscillations are only present for $\mu>1/2$. More precisely, the probability
distribution for the accumulated phase $\Phi_{0}(t)$ can be computed analytically in the diffusion
regime. In this limit $\Phi_0(t)=h N[0,t]$ can be viewed as a positive real number. For $0<\mu<1$, its probability 
distribution is highly non Gaussian and given by
\begin{equation}
P_{0,t}(\phi)=\frac{t}{\mu\,\tau_\phi(h)\,\phi^{1+1/\mu}}
\,L_\mu\left(\frac{t/\tau_\phi(h)}{\phi^{1/\mu}}
\right)
\end{equation}
where $\tau_\phi(h)=\tau_0\,(\Gamma(1-\mu)/h)^{-1/\mu}$ and $L_{\mu}$ denotes the fully asymmetric
L\'evy distribution of index $\mu$. 
For $\mu<1/2$, it decays monotonically whereas for $\mu>1/2$, it grows towards a maximum before decreasing 
rapidely for large values. This maximum can be viewed as a precursor of the maximum 
expected for $1<\mu<2$ close the average value $h t/\langle\tau\rangle$ of the phase (for 
$1<\mu<2$, $P_{0,t}$ is a truncated L\'evy distribution whereas for $\mu\geq 2$ it is Gaussian). 
It is precisely this local maximum that leads to oscillations in the modulus of the dephasing factor.

Note that these oscillations might be related to the Griffith
effect mentioned in ref.~\onlinecite{Schriefl2004}. 

\subsection{The intermediate class $1<\mu<2$}

In the intermediate case of diverging variance, $1<\mu<2$, it is even
harder to find scaling laws analytically. Numerically, one finds
that the decay time of $D_{t_p}(t)$ in the weak coupling regime
scales as $\tau_\phi \propto \tau_{0}/h$.
This leads to a critical coupling $h_{c}(t_{p})$ equal to the r.h.s of \eqref{eq:gc1<mu<2}.
At weak coupling, the decay roughly follows
an exponential $\exp(-t/\tau_\phi)$ up to times of the order of $\tau_\phi$.
At larger times it crosses over to a power law, $D_{t_p}(t) \propto 1/t^\mu$.
Again, at intermediate times and in the weak coupling regime $D_{t_p}(t)$
shows a transcient regime.
Note also that the cross-over between the weak and strong coupling regimes happens
in a much larger ($\mu$-dependent) range of the coupling strength than in
the symmetric case.

\section{Discussion and conclusion}
\label{sec:discussion}

 To conclude, we have proposed a new model for an intermittent classical noise with a $1/f$ power spectrum. 
 Within this model, the noise consists in a succession of pulses, separated by random waiting times. We have shown 
 within this context that the intermittence associated with the $1/f$ characteristics implies a non-stationarity of 
 the noise. Non-stationarity effects are present in some of the regimes of the dephasing of a two level system 
 coupled to this noise as summarized on tables \ref{table:gctauphi} and \ref{table:Dtp} for the symmetric model. 
 In particular, in the strong coupling regime, the dephasing factor 
 decays algebraically in time, with a characteristic time of decay (dephasing time) depending on the 
 age $t_{p}$ of the noise. On the other hand, in the weak coupling regime an initial exponential 
 decay of this dephasing factor is found. However, for low frequency noises and contrarily to the usual behavior, 
 this exponential is stretched and the dephasing time shows an anomalous dependence on the coupling to the noise. After this 
 initial decay, the dephasing factor decays algebraically.
 An important observation is that the critical coupling strength separating the weak and strong coupling regimes does depend 
on the age of the noise when it is non-stationary. More precisely, it decays to zero with the age, such that any 
qubit coupled to a non-stationary noise will eventually fall in the strong coupling regime. 

 One should be careful that the dephasing factor that we studied is defined as a configuration average, or {\it ensemble average} 
 over the noise 
(see eq.~\eqref{eq:defD}). In the usual experimental protocols, information on the quantum statistics of the
qubit is collected in a given sample in successive runs. This corresponds to a {\it time average} in a given configuration. 
These two averages do not coincide in general for non-stationary or aging 
phenomena. Indeed, the non-stationarity of the $1/f$ noise that we considered is closely related to the 
weak ergodicity breaking found in similar trap models for glassy materials\cite{Bouchaud1992}. Thus one should 
be careful in interpreting the non-stationary dephasing factor that we found in some regimes. 
We nevertheless hope that 
the questions raised by our results might lead to some possible experimental setup to better characterize the low frequency 
noise in mesoscopic solids.  


\begin{table*}
\begin{tabular}[t]{|c||c||c|c|}
\hline
Exponent $\mu$ & 
Critical coupling $g_{c}(t_{\mathrm{p}})$ & 
\multicolumn{2}{c|}{Dephasing time $\tau_\phi$} \\
\cline{3-4}
&& 
$g<g_c$ & 
$g>g_c$ \\
\hline \hline
$1<\mu<2$ &
$g_c(t_{\mathrm{p}})^2=\frac{1}{\mu-1}\left[\frac{1}{e}+\left(\frac{\tau_0}{t_{\mathrm{p}}}\right)^{\mu-1}\right]^{1/(\mu-1)}$ & 
$\tau_\phi=\tau_0 g^{-2}$ & 
$\tau_\phi=\tau_0[1/e+(\tau_0/t_{\mathrm{p}})^{\mu-1}]^{-1/(\mu-1)}$ \\
\hline
$ \mu=1$ & $g_c(t_{\mathrm{p}})^2=\frac{\tau_0}{t_{\mathrm{p}}}\ln\left(t_{\mathrm{p}}/\tau_0\right)$ &
$\tau_\phi=\frac{\tau_0}{f_g}|\ln f_g|$ & 
$\tau_\phi=\tau_0 (t_{\mathrm{p}}/\tau_0)^{1-1/e}$    \\
\hline
$ 0<\mu<1$ & 
$g_c(t_{\mathrm{p}})^2\propto\left(\frac{\tau_0}{t_{\mathrm{p}}}\right)^\mu$ &
$\tau_\phi=\tau_0 g^{-2/\mu}$ & 
$\tau_\phi \propto t_{\mathrm{p}}$ \\ 
\hline \hline
\end{tabular}
\caption{\label{table:gctauphi}
Critical coupling and dephasing time in the symmetric model for different waiting time distributions.}
\end{table*}

\begin{table*}
\begin{tabular}[t]{|c||c|c|c||c|c|c|}
\hline
Exponent $\mu$ & 
\multicolumn{3}{c||}{$g>g_\mathrm{c}$} & 
\multicolumn{3}{c|}{$g<g_\mathrm{c}$} \\
\cline{5-7}
& 
\multicolumn{3}{c||}{} & 
\multicolumn{2}{c|}{$t\ll\tau_\phi$} &
 $t\gg\tau_\phi$ \\
\hline \hline 
$1<\mu<2$ &
 \multicolumn{3}{c||}{$\left(\frac{\tau_0}{\tau_0+t}\right)^{\mu-1}-\left(\frac{\tau_0}{\tau_0+t_{\mathrm{p}}+t}\right)^{\mu-1}$} & \multicolumn{2}{c|}{$\exp(-t/\tau_\phi)$} & 
 $t_{\mathrm{p}}<\tau_\phi:\; \exp(-t/\tau_\phi)$ \\
&
\multicolumn{3}{c||}{} &
\multicolumn{2}{c|}{} &
 $t_{\mathrm{p}}>\tau_\phi:\; \left(\frac{\tau_0}{\tau_0+t}\right)^{\mu-1}$\\
\hline \hline
$\mu=1$ &
\multicolumn{3}{c||}{$\frac{1}{\ln(1+t_{\mathrm{p}}/\tau_0)}\ln\left(1+\frac{t_{\mathrm{p}}}{\tau_0+t}\right)$} &
 \multicolumn{2}{c|}{$\exp \left( -f_g \frac{t/\tau_0}{\ln(t/\tau_0)}\right)$}& $1/t$ \\
\hline \hline
$0<\mu<1$ & 
   $t\ll\tau_\phi$  &
    \multicolumn{2}{c||}{$t\gg\tau_\phi$} &
  \multicolumn{2}{c|}{$\mu\simeq1:\;\exp[-(t/\tau_\phi)^{\mu}]$} &
  $\left(\frac{\tau_0}{t}\right)^\mu$ \\
\cline{2-4}
&    $1-A(\mu)(t/t_{\mathrm{p}})^{1-\mu}$ &
\multicolumn{2}{c||}{$(t_{\mathrm{p}}/t)^\mu$} & 
\multicolumn{2}{c|}{$\mu\ll1:\; \frac{1}{1+(t/\tau_\phi)^{\mu}}$} & \\
\hline \hline
\end{tabular}
\caption{\label{table:Dtp}
Summary of the different expressions of the decoherence factor $D_{t_{\mathrm{p}}}$ in the
symmetric model.}
\end{table*}

%
\appendix

\section{Useful results on the sprinkling time distribution}
\label{sec:levy}

\subsection{Definition}
\label{sec:levy:S}

This distribution $S(t)$ is defined as the probability density
that an event occurs exactly at time (date) $t$. Hence $S(t)$
satisfies the equation
\begin{equation}
\label{eq:Int-S} S(t)= \psi(t) + \int_{0}^{t}dt' \psi(t-t') S(t')\,,
\end{equation}
 which states that the spike at $t$ is either the first one, or
follows a previous spike at time $t-t'$. In Laplace transform,
this reads:
\begin{equation}
\label{def-LS} L[S](s) = \frac{L[\psi](s)}{1-L[\psi](s)}\,.
\end{equation}
Whenever $\psi(\tau)$ has a finite mean, $S(t)$ is constant, equal
to $1/\langle \tau \rangle $ with possible large fluctuations ($\mu>1$).

\subsection{Explicit expressions}

 In the case of a Poissonian waiting time distribution $\psi(\tau)=\gamma\,e^{-\gamma \tau}$,
 $S(t)$ is a constant equal to $1/\langle \tau \rangle$ for all times. When $\psi$ has algebraic tails
 at long times, we expect this result to be modified since the rate of events is expected to decrease
 with the sampling of the algebraic tail of $\psi$. 
  
\subsubsection{Case $1<\mu<2$}

The small $s$ expansion of $L[\psi](s)$ is given by:
$$L[\psi](s)\simeq 1 -\ A\tau_0s-\Gamma(1-\mu)\,(s\tau_0)^\mu$$
where $A$ is a numerical constant which depends on the small time
behavior of $\psi(\tau)$. It is given by $A=1/(\mu-1)$ for the
specific case of \eqref{def-P}. Using this form, in the vanishing
$s$ limit, we have:
\begin{equation}
\label{eq:LS:1<mu<2} L[\psi](s)\simeq
\frac{1}{A\tau_{0}s}\,\left(1-\frac{\Gamma(1-\mu)}{A}\,(\tau_{0}s)^{\mu-1}
-A\,\tau_{0}s+\mathcal{O}(s^2)
\right)\,.
\end{equation}
Taking the inverse Laplace transform gives the following
asymptotics for $t\gg \tau_{0}$:
$$S(t)\simeq \langle \tau \rangle^{-1}\,\left(
1+\frac{A^{-1}}{\mu-1}\,\left(\frac{\tau_{0}}{t}\right)^{\mu-1}
\right)\,.$$ 
For the specific case of \eqref{def-P},
\begin{equation}
\label{eq:Sasympt:1<mu<2} S(t)\simeq \langle \tau \rangle^{-1}\,\left(
1+\left(\frac{\tau_{0}}{t}\right)^{\mu-1} \right)\,.
\end{equation}
The sprinkling time distribution decreases algebraically towards
its asymptotic value $S(\infty)=1/\langle \tau \rangle$. This algebraic tail
is the signature of the slow decay of $\psi(\tau)$ for very large
times.

\subsubsection{Case $\mu =1$}

In this case, the Laplace transform of $\psi$ is given by
\begin{equation}
\label{eq:Laplacepsi:mu=1} L[\psi](s)=
1+s\tau_{0}\,e^{s\tau_{0}}\mathrm{Ei}(s\tau_{0})
\end{equation}
where $\mathrm{Ei}$ denotes the exponential integral function which
has the following expansion valid for $x>0$ (eq.8.214 in ref.~\onlinecite{Gradsteyn1980}) :
\begin{equation}
\mathrm{Ei}(-x)=C+\log{(x)}+\sum_{n=1}^{+\infty}
\frac{(-x)^k}{k!\,k}.
\end{equation}
Therefore, keeping only the most singular term in the limit
$s\tau_{0}\rightarrow 0$ we get:
\begin{equation}
L[S](s)=\frac{-1}{s\tau_{0}\,\log{(s\tau_{0})}}.
\end{equation}
The inverse Laplace transform cannot be found exactly but its
asymptotic behavior at large times can be estimated as:
\begin{equation}
S(t)\simeq \frac{1}{\tau_{0}\,\log{(t/\tau_{0})}}.
\end{equation}
In this case, the sprinkling time distribution decays to zero when
$t\rightarrow +\infty$. This means that events are more and more
rare.

\subsubsection{Case $0<\mu < 1$}
In this case, we find
\begin{equation}
\label{eq:LS:mu<1} L[S](s) \simeq_{s\to 0}\frac{(\tau_0s)^{-\mu}}
{\Gamma (1-\mu)}
\end{equation}
which corresponds to
\begin{equation}
\label{eq:Sasympt:mu<1} S(t) \simeq_{t\to \infty } \frac{\sin (\pi
\mu)}{\pi }\frac{1}{\tau_0} \left(\frac{\tau_0}{t}
\right)^{1-\mu}
\end{equation}
In this case also, the sprinkling time distribution decays to zero
when $t\rightarrow +\infty$.

\subsection{Translated sprinkling time distribution}

Let us denote by $S_{t_p}(t)$ the density of events at time
$t_p+t$. By definition $S_{t_p}(t)=S(t_p+t)$. Using the same idea
as above, we can decompose noise histories in two classes: the one
that do not have any event between $t_p$ and $t_p+t$ and the ones
who do have. This leads to an integral equation expressing
$S_{t_p}(t)$ in terms of $\psi$ and $\psi_{t_p}$ defined as the
probability distribution for the time between $t_p$ and the first
event occurring after $t_p$. This integral equation, analogous to
\eqref{eq:Int-S} is:
\begin{equation}
\label{eq:Int-Stp}
S_{t_p}(t)=\psi_{t_p}(t)+\int_0^t\psi(t-\tau)S_{t_p}(\tau)\,d\tau.
\end{equation}
Taking its Laplace transform leads to:
\begin{equation}
\label{eq:LStp} L[S_{t_p}] = \frac{L[\psi_{t_p}]}{1-L[p]}.
\end{equation}

%

\section{Laplace transform and moments of $\psi(\tau),\psi_{t_p}(\tau)$}
\label{sec:renewal}

\subsection{Laplace Transform and moments of $\psi(\tau)$}
\label{sec:renewal:Laplace}

Let us recall some known results on Laplace Transform of algebraic
distributions. The distribution \eqref{def-P} can be expressed in
terms of the reduced variable $x=\tau / \tau_0$:
\begin{equation}\label{def:Px}
\psi(x)= \mu \left(1+x \right)^{-1-\mu}\,.
\end{equation}
It is useful to notice that the Laplace transform of (\ref{def-P})
reads exactly
\begin{equation}
\label{def-LP} L[\psi](s)=\mu  (s\tau_0)^{\mu}
e^{s\tau_0}~\Gamma(-\mu ,s\tau_0)\,.
\end{equation}
Here $\Gamma (-\mu ,s\tau_0)$ is the incomplete Gamma function\cite{Gradsteyn1980}:
$\Gamma(z,\alpha)=\int_{\alpha}^{\infty }e^{-t}t^{z-1}dt$.
The moments of this distribution exist up to order $[\mu]$ where
$[\mu]$ corresponds to the largest integer smaller than $\mu$.
They read:
\begin{align*}
\langle x^{n} \rangle & =
\mu \int_{0}^{\infty} dx~ x^{n} (1+x)^{-1-\mu}\\
&=\mu \int_{0}^{1}dy~ y^{n}(1-y)^{\mu-1-n}\\
&=\mu B(n+1,\mu-n)\equiv \mu \frac{\Gamma(n+1)\Gamma(\mu-n)
}{\Gamma(\mu+1)}
\end{align*}
where $B(x,y)$ is the beta function.

We are interested in the relation between the existence of these
moments, and the small $s$ behavior of the Laplace transform
(\ref{def-LP}).
Expanding the incomplete Gamma function $\Gamma(\mu,s)$ in the
Laplace Transform (\ref{def-LP}) gives:
\begin{align*}
L[\psi](s)&=\mu s^{\mu} e^{s} \left(\Gamma(-\mu)- s^{-\mu}
\sum_{0}^{\infty}\frac{(-1)^{k} s^{k}}{k! (k-\mu)} \right)\\
&= \mu \Gamma(-\mu) s^{\mu}e^{s} +\mu \sum_{n=0}^{\infty}s^{n}
\sum_{k=0}^{n} \frac{(-1)^{n-k}}{k!(n-k)!(\mu+k-n)}
\end{align*}
valid for non-integer $\mu$. The coefficient of $s^n$ in this
expansion reads:
\begin{align*}
\sum_{k=0}^{n} \frac{(-1)^{n+k}}{k!(n-k)!(\mu+k-n)} &=
\sum_{k=0}^{n} \frac{(-1)^{n+k}}{k!(n-k)!}
\int_{0}^{1}dt~ t^{k+\mu-n -1}\\
&=\frac{(-1)^{n}}{n!}\int_{0}^{1}dt~ (1-t)^{n}t^{\mu-n -1}\\
&=\frac{(-1)^{n}}{n!} B(n+1,\mu-n)\,.
\end{align*}
Restoring $\tau_{0}$, we thus get an explicit expression for the
Laplace Transform of $\psi$:
\begin{equation}
\label{eq:expansionLP} L[\psi](s)= \mu \Gamma(-\mu)
(s\tau_{0})^{\mu}e^{s}
+\sum_{n=0}^{\infty}\frac{(-1)^{n}}{n!}(s\tau_{0})^{n}
\langle x^{n}\rangle
\end{equation}
The terms $n<\mu$ of the last expansion corresponds to the moments
of $\psi(x)$, and we recover the expected results: the small $s$
expansion of $L[\psi](s)$ consist in a integer function that gives
the existing moments of $\psi$ and a second part that is specific
of the tails of $\psi(x)$ if it decays more slowly than an
exponential (algebraic tails).

\subsection{Double Laplace transform of $\psi_{t_{p}}(\tau_{1})$}

In Ref.~\onlinecite{Godreche2001}, Godr{\`e}che and Luck used a
direct CTRW method to derive the double Laplace transform of
$\psi_{t_p}(\tau_{1})$. Their results can be recovered
straightforwardly from the renewal equation
\eqref{eq:integralforpsitp}. We will use the following notation
for the double Laplace transform
\begin{multline}
\label{eq:dbleLaplace}
L_{t_{p},\tau_{1}}[\psi_{t_{p}}(\tau_{1})](u,s) = \\
\int_{0}^{+\infty} d\tau_{1}~ e^{-s\tau_{1}}\int_{0}^{+\infty}
dt_{p}~ e^{-ut_{p}} ~\psi_{t_{p}}(\tau_{1})
\end{multline}
Let us first focus on the $t_{p}$-Laplace transform of the
integral in eq. \eqref{eq:integralforpsitp}. Changing the
integration variable from $t_{p}$ to $t'=t_{p}-\tau_{1}$, it reads
\begin{multline}
\int_{0}^{+\infty} dt_{p}~ e^{-ut_{p}}
 \int_0^{t_p}d\tau~ S(t_p-\tau) \psi(\tau_{1}+\tau)\\
=L[S](u)\int_{0}^{\infty}d\tau e^{-u\tau}\psi(\tau_{1}+\tau)
\end{multline}
This provides the following expression for the double Laplace
transform of $\psi_{t_{p}}(\tau_{1})$ :
\begin{align}\nonumber
L[\psi_{t_{p}}](u,s)  &=(1+L[S](u))~
L_{t_{p},\tau_{1}}[\psi(t_{p}+\tau_{1})](u,s)\\
\label{eq:Laplaceh} &=\frac{L_{t_{p},\tau_{1}}
[\psi(t_{p}+\tau_{1})](u,s)}{1-L[\psi](u)}
\end{align}
The final result can be obtained by considering
 the double Laplace transform of $\psi(\tau_{1}+t_{p})$ :
using $\int_{\tau_1}^{+\infty}\psi(\tau)e^{-s\tau}d\tau=
L[\psi](s) - \int_0^\tau \psi(\tau)e^{-s\tau}d\tau$, we obtain
(with $\tau=\tau_{1}+t_{p}$)
\begin{align}
 L_{t_{p},\tau_{1}}&[\psi(t_{p}+\tau_{1})](u,s)\nonumber \\\nonumber
&= \int_{0}^{+\infty}d\tau_{1}e^{-(s-u)\tau_{1}}
\int_{\tau_{1}}^{\infty}d\tau~e^{-u\tau}\psi(\tau)\\\nonumber &=
\frac{L[\psi](u)}{s-u}-\int_{0}^{\infty}d\tau_1\int_{\tau_1}^{\infty}d\tau
e^{-(s-u)\tau_{1}}
e^{-u\tau}\psi(\tau) \\
\label{eq:dbleLaplaceP-t-tp}
&=\frac{1}{s-u}\left(L[\psi](u)-L[\psi](s) \right)
\end{align}
Plugging \eqref{eq:dbleLaplaceP-t-tp} into \eqref{eq:Laplaceh}, we
obtain\cite{Godreche2001}:
\begin{equation}
\label{eq:psi-tp-dbleLaplace} L[\psi_{t_{p}}](u,s) = \frac{1}{s-u}
\frac{L[\psi](u)-L[\psi](s)}{1-L[\psi](u)}
\end{equation}
This equation can be used to infer explicit expressions for
$\psi_{t_p}(t)$ by performing the appropriate inverse Laplace
transforms (see appendix \ref{sec:explicitpsitp}).

\subsection{Moments of $\psi_{t_{p}}(\tau)$}

The expansion of $L[\psi](s)$ in powers of $s$ can be used to extract the
long $t_{p}$ behavior of the moments of $\psi_{t_{p}}(\tau)$.
 For $\mu >1$, we expect $\psi_{t_p}$ to
have the same algebraic decay than $\psi$ at infinity. Therefore,
only the first $[\mu]$ moments of $\psi_{t_p}$ are expected to
exist. We will expand the double Laplace transform
(\ref{eq:psi-tp-dbleLaplace}) of $L[\psi_{t_p}](u,s)$ into powers
of $s$ for small values of $u$. The first coefficients of the $s$
expansion corresponds to the Laplace transform with respect to
$t_{p}$ of the moments of $\psi_{t_{p}}(\tau)$  which we denote here by
$T_n(t_p)=\langle\tau_1^{n}\rangle_{\psi_{t_p}}$. When
considering eq.~\eqref{eq:psi-tp-dbleLaplace}, two different
contributions from $(L[\psi](u)-L[\psi](s))/(s-u)$ appear. The non
integer powers can be expressed as
\begin{equation}
\frac{u^\mu-s^\mu}{u-s} = \sum_{k=0}^\infty s^k\,u^{\mu-1-k}
-\sum_{k=0}^\infty s^{\mu+k}\,u^{-k-1},
\end{equation}
while integer powers give a finite sum:
\begin{equation}
\frac{u^n-s^n}{u-s} = \sum_{k=0}^{n-1}s^k\,u^{n-1-k}.
\end{equation}
Note that fractional powers are of the form $s^{\mu+k}$ where
$k\geq 0$, which confirms that only the first $[\mu]$ moments
exist. Assuming the following notation:
\begin{equation}
L[\psi](s)=\sum_{n=0}^{[\mu]}
\frac{(-s)^n}{n!}\,\langle\tau^n\rangle -\frac{\tau_{\mu}}{\Gamma(\mu)}\,s^\mu\,,
\end{equation}
the coefficient of the $s^N$ term in the expansion
of \eqref{eq:psi-tp-dbleLaplace} in powers of $s$ is given by:
\begin{multline*}
\frac{1}{1-L[\psi](u)}\,\left(\frac{\tau_\mu}{\Gamma(\mu)}\,u^{\mu-1-N}
\right.\\
\left.
-\sum_{m=N+1}^{[\mu]}\frac{\langle \tau^m \rangle}{m!}\,(-1)^m\,u^{m-1-N}\right).
\end{multline*}
Note that the second term is present only for $N\leq [\mu]-1$.
Assuming that $\mu>1$, $1-L[\psi](u)$ can safely be replaced by
$\langle\tau\rangle u$ in order to extract the low $u$ behavior of the
above expression. From this, we infer the Laplace transform of
$T_N(t_p)$:
\begin{multline}
L_{t_{p}}[T_N](u)  = 
(-1)^{N}\,\frac{N!\,\tau_\mu}{\Gamma(\mu)\,\langle\tau\rangle}\,
u^{\mu-2-N}\\
\label{eq:moments2}
  -   \sum_{m=-1}^{[\mu]-m-2} \frac{(-1)^m(m+1)!}{(N+m)!}\,
 \frac{\langle \tau^{N+m} \rangle}{\langle \tau \rangle}\,u^m.
\end{multline}
This formula shows that for $N<[\mu]$, there is a limiting value
for $t_p$ going to infinity, given by the $u^{-1}$ term in the sum
\eqref{eq:moments2}. Regular terms that
appear in the sum \eqref{eq:moments2} contain the short time
behavior in $t_p$ and, in the large $t_{p}$ limit, the non integrer power in the r.h.s gives 
an algebraic decaying contribution. The limiting value of $T_{N}(t_p)$ for $N\leq [\mu]-1$ is given
by:
\begin{equation}
\label{eq:momentlimit}
T_N(\infty)=\frac{\langle \tau^{N+1}\rangle}{(N+1)\langle \tau \rangle}.
\end{equation}
This result coincides with \eqref{eq:tau1:mu>2} found previously
for $N=1$ as soon as $\mu>1$.
For $N=[\mu]$, the regular contribution to \eqref{eq:moments2} is not there
anymore. The $[\mu]$th moment has an algebraic sublinear
dependance in $t_p$:
\begin{equation}
\label{eq:ptp:moments:algebraicone}
T_{[\mu]}(t_p)\sim
\frac{\tau_\mu}{\langle \tau \rangle}\,t_p^{1-(\mu-[\mu])}.
\end{equation}
To summarize, the algebraic tail of $\psi(\tau)$ contaminates
$\psi_{t_p}$ not only through the divergence of its high moments
but also through algebraic corrections of the lower ones and the
sublinear algebraic behavior of the last finite one.

In particular these results imply:
\begin{itemize}
\item For $1<\mu<2$, the first moment $\langle \tau_{1} \rangle(t_p)$ increases
sublinearily as obtained from a direct computation.
\item For $2<\mu<3$, the second moment also increases sublinearily
although it is finite for any $t_p$. Only when $\mu>3$ do we have
finite limits for both the first and second moments of
$\psi_{t_p}$ in the limit $t_p$ going to infinity.
\end{itemize}

%
\section{Explicit expressions for $\psi_{t_p}(\tau_1)$}
\label{sec:explicitpsitp}

\subsection{Expression of $\psi_{t_{p}}(\tau_{1})$ for $1<\mu <2$}
\label{sec:explicitpsitp:1<mu<2}

The small $s$ expansion of $L[\psi](s)$ reads
$L[\psi](s)=1-s\langle \tau \rangle+ \mu \Gamma(-\mu)(s\tau_0)^{\mu
}$ (we have to keep all the terms up to the first non-integer
power). Once again, we set $\tau_{0}=1$ for simplicity.
Plugging it into (\ref{eq:psi-tp-dbleLaplace}) gives:
\begin{align}\nonumber
L[\psi_{t_{p}}](u,s) &= \left(\langle \tau \rangle -\mu
\Gamma(-\mu)\frac{s^{\mu }-u^{\mu}}{s-u} \right)
\frac{1}{u\langle \tau \rangle-\mu \Gamma(-\mu)u^{\mu} }\\
&=\left(\frac{1}{s}- A~ \frac{s^{\mu }-u^{\mu}}{s(s-u)}\right)
\left(1+A~ s^{\mu-1 }\right)
\end{align}
where $\tau_{0}$ has been set to one for simplicity and $A=\mu
\Gamma(-\mu)/\langle \tau \rangle=-\Gamma(1-\mu)/\langle \tau \rangle
  =\Gamma(2-\mu)$ (we used $\langle \tau \rangle=\tau_{0}/(\mu-1)
$). Doing the inverse Laplace transform over $s$ yields
\begin{align}\label{eq:laplaceh1<mu<2}
&L[\psi_{t_{p}}](u) =\left(
1+\frac{A~t_{p}^{1-\mu}}{\Gamma(2-\mu)} \right)
\\\nonumber
&+A~u^{\mu-1} \left(e^{u t_{p}}\frac{\Gamma(1-\mu
,ut_{p})}{\Gamma(1-\mu)} -1
 \right)
+O(u^{2},u^{2(\mu-1)})
\end{align}
Note that the first constant reads
\begin{equation}
1+\frac{A~t_{p}^{1-\mu}}{\Gamma(2-\mu)} = 1+\frac{1}{\tau_0}
\left(\frac{\tau_0}{t_{p}} \right)^{\mu-1}
\end{equation}
Restoring $\tau_{0}$, the inverse Laplace transform over $u$ gives:
\begin{equation}
\label{eq:psitp:1<mu<2} \psi_{t_{p}}(\tau_{1})
=\frac{1}{\langle \tau \rangle } \left(\left(\frac{\tau_0}{\tau_{1}}
\right)^{\mu}- \left(\frac{\tau_0}{\tau_{1}+t_{p}}
\right)^{\mu}
 \right)\,.
\end{equation}

\subsection{Expression of $\psi_{t_{p}}(\tau_{1})$ for $0<\mu <1$}
\label{sec:explicitpsitp:mu<1}

Let us use the small $s$ expansion (\ref{eq:expansionLP}) of
$L[\psi](s)$ : $L[\psi](s)=1+\mu \Gamma(-\mu)(s\tau_0)^{\mu }$.
Using the double Laplace transform \eqref{eq:psi-tp-dbleLaplace}
and setting $\tau_{0}=1$ for simplicity, we get:
\begin{equation}
L[\psi_{t_{p}}](s,u) =\frac{1}{u-s} \left(1-\left(\frac{u}{s}
\right)^{\mu } \right)
\end{equation}
The only assumption in deriving this expression is that both
$t_{p}$ and $\tau_{1}$ are large compared to $\tau_0$ (we
retained only the first terms of the $s$ and $u$ expansions). This
expression can be exactly Laplace inverted in $s$ :
\begin{equation}
\label{eq:psitp:0<mu<1:Laplace} L[\psi_{t_{p}}](u) =
e^{ut_{p}}\frac{\Gamma(\mu ,st_{p})}{\Gamma(\mu )}.
\end{equation}
We can now perform the inverse Laplace transform over $u$ to
obtain:
\begin{equation}
\label{eq:psitp:0<mu<1} \psi_{t_p}(\tau_{1})=
\frac{\sin\pi\mu}{\pi} \; \frac{1}{\tau_{1}+t_p}
\left(\frac{t_p}{\tau_{1}}\right)^\mu.
\end{equation}
Note that in this case, the short time scale $\tau_{0}$ does not
appear in $\psi_{t_p}(\tau_1)$ contrarily to the cases $\mu>1$.
For $\tau_{1}\gg t_{p}$, this distribution behaves like
\begin{equation}
\psi_{t_p}(\tau_{1}) \simeq \frac{\sin (\pi \mu ) }{\pi \mu }
\left(\frac{t_{p}}{\tau_0} \right)^{\mu}\psi(\tau_{1})\,,
\end{equation}
which shows that for very large waiting times, we have the same
algebraic tail than $\psi$ up to a $t_p$ dependent renormalization
factor. For $\tau_{1}\ll t_{p}$, we get another algebraic tail :
\begin{equation}
\psi_{t_p}(\tau_{1}) \simeq \frac{\sin (\pi \mu ) }{\pi
}\frac{1}{t_{p}} \left(\frac{t_{p}}{\tau_{1}}\right)^{\mu}\,.
\end{equation}

\subsection{Expression of $\psi_{t_{p}}(\tau_{1})$ for $\mu =1$}
\label{sec:explicitpsitp:mu=1}

In the marginal case $\mu = 1$, because of its logarithmic
variation, $S(t_p-\tau)$ can be replaced by $\frac{1}{\tau_{0}
\log(t_p/\tau_{0})}$ in the integral equation
\eqref{eq:integralforpsitp}. This approximation leads to:
\begin{multline}
\label{eq:htpu=1} \psi_{t_p}(\tau_{1}) \approx \psi(\tau_{1}+t_p)\\
+ \frac{1}{\log(t_p/\tau_{0})}
\frac{1}{\tau_{1}+t_p+\tau_{0}}\frac{t_p}{\tau_{1}+\tau_{0}}.
\end{multline}
In the following we always consider situations in which $t_p,
\tau_{1} \gg \tau_{0}$, for which the first term in
\eqref{eq:htpu=1} can be neglected. Therefore, we get the
following asymptotics:
\begin{equation}
\label{eq:psitp:mu=1} \psi_{t_p}(\tau_{1})\simeq
\frac{1}{\log(t_p/\tau_{0})}
\frac{1}{\tau_{1}+t_p}\frac{t_p}{\tau_{1}}.
\end{equation}
Note that in this case, the average $\langle \tau_{1} \rangle_{\psi_{t_{p}}}$ is infinite.

%
\section{Analytic results on $\Pi_0$}
\label{sec:Pi0}

Explicit expressions for $\Pi_0(t_p,t)$ immediately follow from
its definition and from explicit expressions for $\psi_{t_p}(t)$
in the limit $t\gg \tau_{0}$.
In the case $0<\mu<1$, the result of integration is an
hypergeometric function\cite{Gradsteyn1980} :
\begin{equation}
\label{eq:Pi0:0<mu<1} \Pi_0(t_p,t) =
\frac{\sin{(\pi\mu)}}{\pi\mu}\,\left(
\frac{t_p}{t}\right)^\mu\,{}_2F_1(\mu,1;1+\mu;-\frac{t_p}{t})\,.
\end{equation}
In this case, $\Pi_0$ does not depend anymore on the cutoff
$\tau_{0}$ and exhibits an aging behavior since it only depends on the
ratio of time scales $t/t_p$. Using the inversion properties of
hypergeometric function \cite[eq.~9.131.1]{Gradsteyn1980}, an alternative
expression can be obtained:
\begin{equation}
\label{eq:Pi0:0<mu<1:v2} \Pi_0(t_p,t) =
\frac{\sin{(\pi\mu)}}{\pi\mu}\,\left(
\frac{t_p}{t_p+t}\right)^\mu\,{}_2F_1(\mu,\mu;1+\mu;\frac{t_p}{t_p+t})\,.
\end{equation}
Equation \eqref{eq:Pi0:0<mu<1} is well suited to the limit $t\ll
t_p$ whereas eq.~\eqref{eq:Pi0:0<mu<1:v2} is better suited to the
study of the $t_p\ll t$ limit.

In the case $1<\mu<2$, the final result still depends on $\tau_{0}$.
Using eq. \eqref{eq:psitp:1<mu<2}, we get:
\begin{equation}
\Pi_0(t_p,t)\simeq\left(\frac{\tau_{0}}{t}\right)^\mu
-\left(\frac{\tau_{0}}{t_p+t}\right)^\mu
\end{equation}
In the limit of vanishing $\tau_{0}$,  $\Pi_0(t_p,t)$ vanishes. Hence in this 
limit, on times scales long compared to
$\bar{\tau}$, there are always somes events in time intervals of duration
$t$ contrarily to the case where $\mu<1$.

In the $\mu=1$ case, using eq.~\eqref{eq:psitp:mu=1}, we get:
\begin{equation}
\Pi_0(t_,t)\simeq
\frac{\log{\left(1+\frac{t_p}{t}\right)}}{\log{(t_p/\tau_{0})}}\,,
\end{equation}
which shares some characteristics with the two previous cases. On
one hand, as in the $1<\mu<2$ case, it still decays to zero at
fixed $t_p$ and $t$ in the limit $\tau_{0}\rightarrow 0$. On the
other hand, exactly as in the $0<\mu<1$ case, the limits
$t/t_p\rightarrow 0$ and $t/t_p\rightarrow 1$ satisfy:
\begin{eqnarray}
\lim_{t\rightarrow \infty}\Pi_0(t_p,t) & = & 0\,,\\
\lim_{t_p\rightarrow \infty}\Pi_0(t_p,t) & = & 1\,.
\end{eqnarray}


\end{document}